\immediate\write18{makeindex \jobname.nlo -s nomencl.ist -o \jobname.nls}

\documentclass[preprint,5p,twocolumn]{elsarticle}
\usepackage{amsmath}
\usepackage{booktabs}
\usepackage{array}
\usepackage{xcolor}
\usepackage[noabbrev,capitalise]{cleveref}
\usepackage{enumitem}
\usepackage{xfrac}
\usepackage{siunitx}
\usepackage{graphicx}
\usepackage{subcaption}
\usepackage[hyphens]{url}
\usepackage{nomencl}
\usepackage{framed}
\usepackage{multicol}
\usepackage{fancyhdr}

\makenomenclature
\renewcommand*\nompreamble{\begin{multicols}{2}}
\renewcommand*\nompostamble{\end{multicols}}

\biboptions{square,comma,numbers,sort&compress}

\begin{document}

\journal{Progress in Aerospace Sciences}

\begin{frontmatter}

\title{The Benefits of Very Low Earth Orbit for Earth Observation Missions}

\author[UoM]{N.H.~Crisp\corref{cor1}}
\ead{nicholas.crisp@manchester.ac.uk}

\author[UoM]{P.C.E.~Roberts}
\author[UoM]{S.~Livadiotti}
\author[UoM]{V.T.A.~Oiko}

\author[UoM]{S.~Edmondson}
\author[UoM]{S.J.~Haigh}
\author[UoM]{C.~Huyton}
\author[UoM]{L.~Sinpetru}
\author[UoM]{K.L.~Smith}
\author[UoM]{S.D.~Worrall}

\author[Deimos]{J.~Becedas}
\author[Deimos]{R.M.~Dom\'{i}nguez}
\author[Deimos]{D.~Gonz\'{a}lez}

\author[Gomspace]{V.~Hanessian}
\author[Gomspace]{A.~M{\o}lgaard}
\author[Gomspace]{J.~Nielsen}
\author[Gomspace]{M.~Bisgaard}

\author[Stuttgart]{Y.-A.~Chan}
\author[Stuttgart]{S.~Fasoulas}
\author[Stuttgart]{G.H.~Herdrich}
\author[Stuttgart]{F.~Romano}
\author[Stuttgart]{C.~Traub}

\author[UPC]{D.~Garc\'{i}a-Almi\~{n}ana}
\author[UPC]{S.~Rodr\'{i}guez-Donaire}
\author[UPC]{M.~Sureda}

\author[MSSL]{D.~Kataria}

\author[CNU]{R.~Outlaw}

\author[Euroconsult]{B.~Belkouchi}
\author[Euroconsult]{A.~Conte}
\author[Euroconsult]{J.S.~Perez}
\author[Euroconsult]{R.~Villain}

\author[concentris]{B.~Hei{\ss}erer}
\author[concentris]{A.~Schwalber}

\cortext[cor1]{Corresponding author.}

\address[UoM]{The University of Manchester, Oxford Road, Manchester, M13~9PL, United Kingdom}
\address[Deimos]{Elecnor Deimos Satellite Systems, Calle Francia 9, 13500 Puertollano, Spain}
\address[Gomspace]{GomSpace A/S, Langagervej~6, 9220 Aalborg East, Denmark}
\address[Stuttgart]{Institute of Space Systems (IRS), University of Stuttgart, Pfaffenwaldring~29, 70569 Stuttgart, Germany}
\address[UPC]{UPC-BarcelonaTECH, Carrer de Colom~11, 08222 Terrassa, Barcelona, Spain}
\address[MSSL]{Mullard Space Science Laboratory (UCL), Holmbury St. Mary, Dorking, RH5~6NT, United Kingdom}
\address[CNU]{Christopher Newport University, Newport News, Virginia 23606, USA}
\address[Euroconsult]{Euroconsult, 86~Boulevard de Sébastopol, 75003 Paris, France}
\address[concentris]{concentris research management gmbh, Ludwigstraße~4, D-82256 F\"{u}rstenfeldbruck, Germany}

\begin{abstract}
Very low Earth orbits (VLEO), typically classified as orbits below approximately \SI{450}{\kilo\meter} in altitude, have the potential to provide significant benefits to spacecraft over those that operate in higher altitude orbits. This paper provides a comprehensive review and analysis of these benefits to spacecraft operations in VLEO, with parametric investigation of those which apply specifically to Earth observation missions. The most significant benefit for optical imaging systems is that a reduction in orbital altitude improves spatial resolution for a similar payload specification. Alternatively mass and volume savings can be made whilst maintaining a given performance. Similarly, for radar and lidar systems, the signal-to-noise ratio can be improved. Additional benefits include improved geospatial position accuracy, improvements in communications link-budgets, and greater launch vehicle insertion capability. The collision risk with orbital debris and radiation environment can be shown to be improved in lower altitude orbits, whilst compliance with IADC guidelines for spacecraft post-mission lifetime and deorbit is also assisted. Finally, VLEO offers opportunities to exploit novel atmosphere-breathing electric propulsion systems and aerodynamic attitude and orbit control methods.

However, key challenges associated with our understanding of the lower thermosphere, aerodynamic drag, the requirement to provide a meaningful orbital lifetime whilst minimising spacecraft mass and complexity, and atomic oxygen erosion still require further research. Given the scope for significant commercial, societal, and environmental impact which can be realised with higher performing Earth observation platforms, renewed research efforts to address the challenges associated with VLEO operations are required.
\end{abstract}


\end{frontmatter}

\section{Introduction} \label{S:Introduction}
Earth observation (EO) spacecraft and space-systems provide imagery and other remote-sensing data types which are being used for an increasing number of important applications with global significance through industrial, economic, societal, and environmental impacts. Common applications of EO data include environmental monitoring, maritime surveillance, intelligence and homeland security, land management and agriculture, meteorology, and disaster monitoring and response management \cite{Kansakar2016}. As a result of this broad range of applications and global reach, EO from space has been recognised by the United Nations as having a key contributing role towards the achievement of their 17 Sustainable Development Goals \cite{UN2018}.

Operation of spacecraft at lower altitude orbits can be linked to a number of benefits which are particularly relevant for EO applications which profit from global coverage without the inherent constraints of airspace restrictions and limited range and duration. However, at present, few vehicles operate sustainably and for useful durations in the altitude range between high reconnaissance aircraft at \SI{26}{\kilo\meter} (eg. the SR71 Blackbird) and the lowest space platforms at around \SI{450}{\kilo\meter}. These orbits have generally been avoided due to the high cost of spacecraft development, launch, and challenges associated with atmospheric drag which either necessitates the use of a capable propulsion system or significantly limits the mission lifetime.

Despite these challenges, there are a few notable classes of spacecraft which have done, and continue to operate in this altitude range. Military reconnaissance spacecraft, for example in the early DISCOVERER/CORONA satellite programme \cite{Clausen2012}, tolerated very short mission lifetimes in VLEO to provide high resolution surveillance imagery. More recent Keyhole satellites have utilised eccentric orbits with low perigees ($<$\SI{300}{\kilo\meter}) \cite{Richelson1984,Wright2005} to provide longer mission lifetimes, but are therefore also limited in their imaging operations due to the orbit eccentricity. Other missions, including scientific spacecraft (eg. GOCE \cite{Drinkwater2007}) or orbital shuttles (eg. Space Shuttle, X-37) have utilised highly-capable propulsion systems to enable longer duration activities, but are not commercially viable for EO applications. Finally, space stations, most notably the International Space Station (ISS) and previously Mir orbit below \SI{450}{\kilo\meter}, but require resupply missions to provide propellant for orbit maintenance.

Recent technology development, in particular component and subsystem miniaturisation, has enabled significant cost-reduction and enabled more agile spacecraft development cycles. This re-evaluation of traditional spacecraft development has also lead to the establishment of the emerging commercial, so-called ``NewSpace'' industry \cite{Sweeting2018}. Concern about the increasing debris population in higher orbital ranges has also called for mitigation measures and alternative approaches for ongoing spacecraft operation in LEO \cite{Klinkrad2013}. With the introduction of frequent and affordable orbital insertion opportunities (in comparison to dedicated launch) from the ISS and the promise of new commercial launch vehicles, very low Earth orbits (VLEOs) have recently become an attractive proposition. Commercial exploitation of these lower orbital altitudes has already begun, for example by the Planet Labs Flock and Spire Global Lemur-2 CubeSat constellations \cite{Bandyopadhyay2016,Wekerle2017}.  

The VLEO altitude range is principally characterised by the presence of aerodynamic forces which can have a significant effect on the orbital and attitude dynamics of a spacecraft. A nominal altitude of \SIrange{450}{500}{\kilo\meter} is typically applied as the upper threshold for VLEO \cite{VirgiliLlop2014a,Roberts2017}, but this is in reality dependent on the atmospheric conditions and can vary significantly with the solar cycle. The term Super Low Earth Orbit (SLEO) has also been applied, albeit less widely, to orbits with a perigee below \SI{300}{\kilo\meter} \cite{Fujita2008}.

Current research related to VLEO spacecraft principally seeks to enable sustained operations at these lower altitudes, for example through the identification and characterisation of low-drag materials and surface coatings, development of aerodynamic attitude and orbit control, and design of propulsive drag compensation \cite{Fujita2008,Roberts2017,Romano2018c}. 

Through these developments, it is hoped that the cost of launching to and operating in VLEO can be significantly reduced whilst maintaining or improving the resolution and quality of data products. This will consequently improve the downstream cost and availability of imagery and data used in programmes such as maritime surveillance, intelligence and security, land management, precision agriculture and food security, and disaster monitoring with potentially high humanitarian, societal, and commercial impact.

Given the mounting interest in operating spacecraft in the VLEO altitude range, the aim of this paper is to provide a comprehensive overview and analysis of the wide-ranging benefits this non-traditional regime offers, both with respect to general spacecraft operations and specifically to EO applications.

\section{Benefits of Very Low Earth Orbits}
Several studies have discussed various benefits of operating satellites, principally for Earth observation applications in lower altitude regimes. \citet{Wertz2012a} focus on moderately elliptical orbits with perigees and apogees generally below \SI{300}{\kilo\meter} and \SI{500}{\kilo\meter} respectively with the aim of avoiding the build-up of orbital debris. \citet{Shao2014} consider the design of small satellites operating at lower altitudes using a performance-based cost modelling approach. This study demonstrates the cost-benefit of operating at lower altitudes for both improved coverage and resolution, principally due to a reduction in necessary size and mass of the payload. Furthermore, it is noted that these lower altitude systems carry a lower mission risk (from production and launch failures) due to deployment in greater numbers. \citet{RamioTomas2014} consider the top-level parametric design of synthetic aperture radar (SAR) spacecraft, demonstrating that their operation in VLEO is feasible and may offer advantages in cost and revisit when operated in constellations.

This review builds upon the initial collection of benefits of VLEO identified by \citet{Eves2013} and \citet{VirgiliLlop2014a}, providing further and more indepth analysis. First, a comprehensive review and analysis of the benefits of orbital altitude reduction on spacecraft operations with a focus on EO applications is presented. The different modes of Earth observation (optical, radar, and infrared systems) are then considered and the variation in system performance with orbital altitude is analysed.

\subsection{Orbit Geometry}

\begin{figure}
	\centering
	\def\svgwidth{\linewidth}
	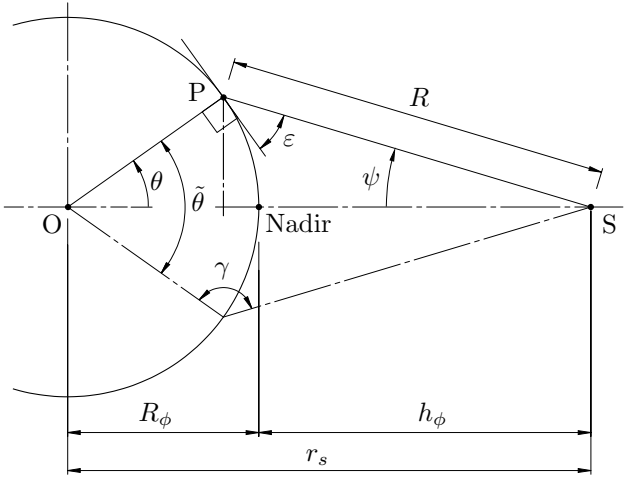
	\caption{Geometry of satellite sensor field of regard. Adapted from \citet{Vallado2013}.}
	\label{F:FootprintGeometry}
\end{figure}

The coverage of a spacecraft in orbit can be defined by the angular field of regard $\psi$ or total footprint area which is available to a given sensor. These parameters and the associated geometry are described in \cref{F:FootprintGeometry}. Alternatively, the instantaneous view of the sensor can be considered, yielding the field of view or sensor footprint area which fall within the angular field of regard.

For a given angular field of regard $\psi$, the central angle $\theta$ from the centre of an assumed circular Earth can be calculated from the orbital altitude via the slant range $R$ and an intermediate angle $\gamma$ \cite{Vallado2013}.
\begin{equation}
\gamma = \sin^{-1} \frac{r_s \sin{\psi}}{R_{\phi}}
\end{equation}
\begin{equation}
R = R_{\phi}\cos{\gamma} + r_s\cos{\psi}
\end{equation}
\begin{equation}
\theta = \sin^{-1} \left( \frac{R\sin{\psi}}{R_\phi} \right)
\end{equation}

The circular footprint area $A_F$ projected on the ground at the nadir can then be approximated using solid angles.
\begin{equation}
\label{E:FootprintArea}
A_F = 2\pi (1-\cos{\theta}) R_{\phi}^{2}
\end{equation}

For a fixed angular field of regard, the footprint area and therefore total available coverage decreases with reducing orbital altitude. However, a number of benefits can be associated with reducing the range to the Earth's surface and will be discussed in the following sections.

\subsection{Spatial Resolution}
\begin{figure} 
	\centering
	\begin{subfigure}[b]{0.49\linewidth}
	\centering
	\includegraphics[width=\linewidth,trim={30 30 30 30},clip]{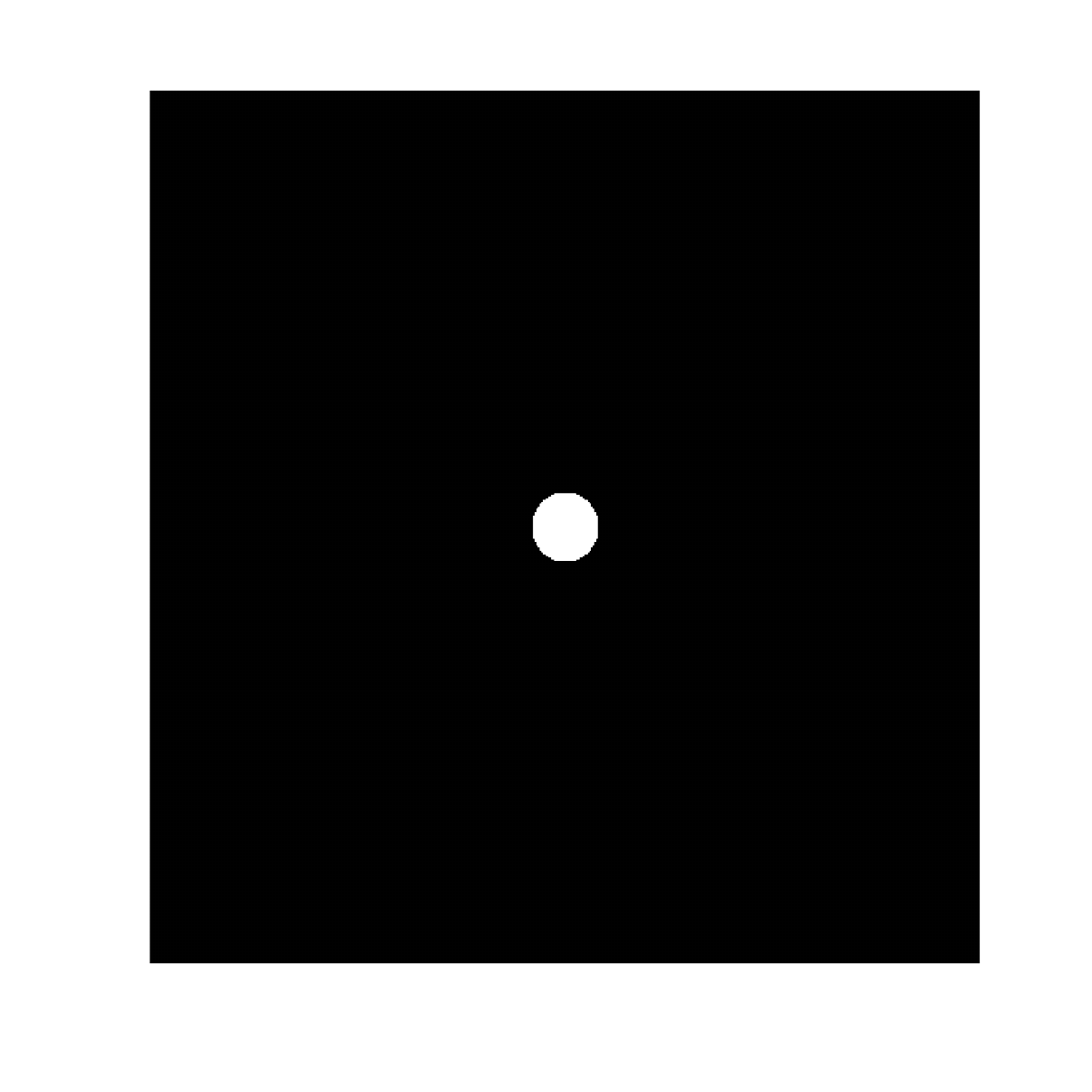}
	\caption{Input Source.}
	\end{subfigure}
	\hfill
	\begin{subfigure}[b]{0.49\linewidth}
	\centering
	\includegraphics[width=\linewidth,trim={30 30 30 30},clip]{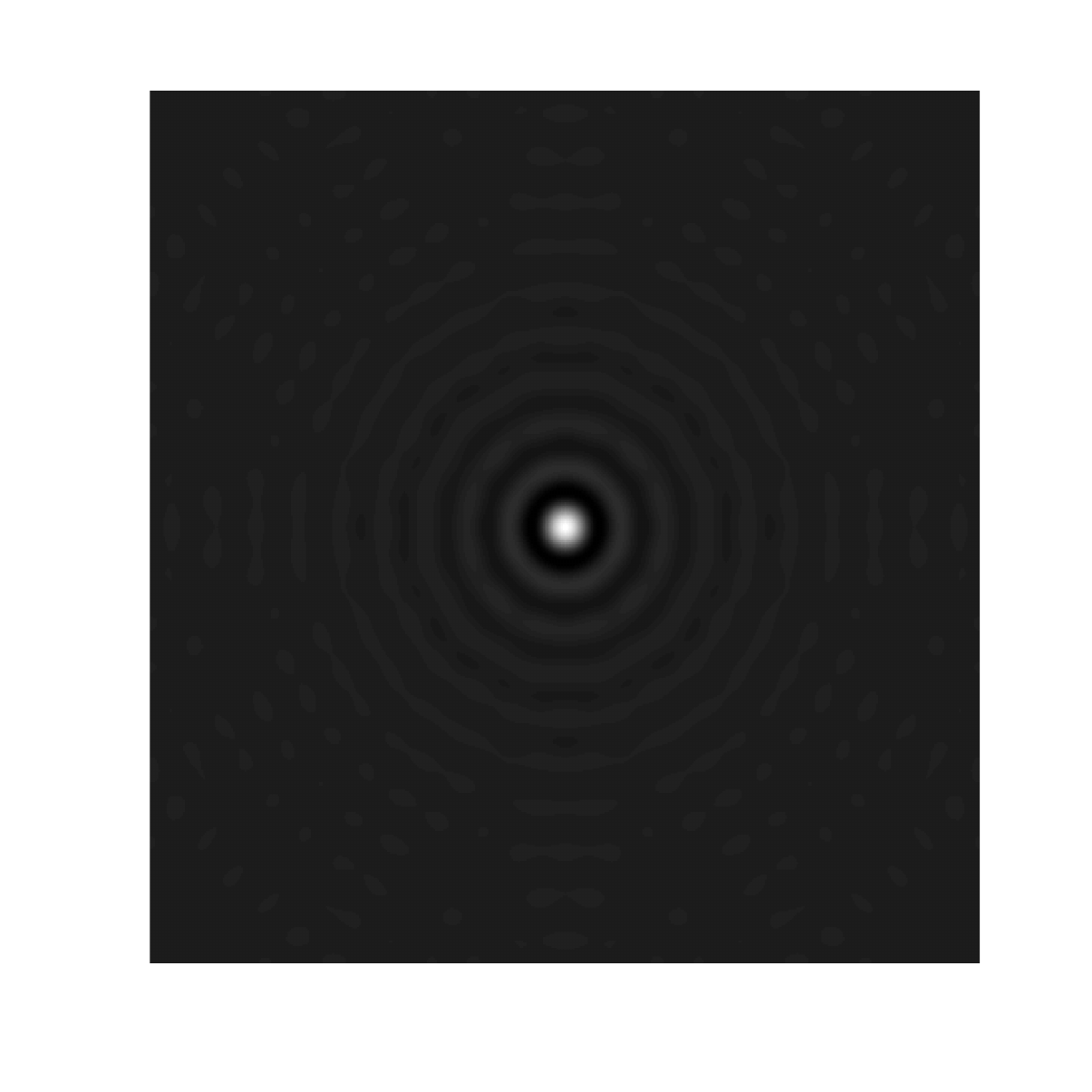}
	\caption{Output Pattern.}
	\end{subfigure}
	\caption{Computer generated demonstration of the point spread function of an optical aperture with a perfect lens (ie. diffraction limited).}
	\label{F:PSF}
	\end{figure}

Due to aberrations, diffraction, and distortions, the imaging of a point-source of light through an optical aperture, projected onto an image plane becomes blurred and can be described by a point-spread function (PSF). Due to the interference of the light, a diffraction pattern (an Airy disk) surrounding the central point can be observed, shown in \cref{F:PSF}, even through perfectly constructed lenses. The quality of an optical lens will further affect the PSF and therefore the clarity of the collected image.

The angular or spatial resolution describes the ability of an imaging device to distinguish between individual points with small separation rather than seeing a single combined or convoluted image. For an imaging system, the theoretical maximum resolution is constrained only by the diffraction of the light or radiated beam through the lens or antenna, described by the Rayleigh criterion which is based on the diffraction-limited PSF (Airy disk) of two neighbouring points \cite{Schott2007}.
\begin{equation}
\sin{\delta_\Theta} = 1.22 \frac{\lambda}{D} 
\end{equation}
Here $\delta_\Theta$ is the angular resolution in radians, $\lambda$ the wavelength, and $D$ the diameter of the lens or antenna aperture.

The corresponding diffraction limited resolution, or ground resolution distance ($\mathit{GRD}$), can be obtained from the angular resolution by simple trigonometry using the range $R$ and the small-angle approximation.
\begin{equation}
\delta_\Theta = \sin^{-1} 1.22 \frac{\lambda}{D} \approx \tan^{-1}(\frac{\mathit{GRD}}{R})
\end{equation}
\begin{equation}
\label{E:GRD}
\mathit{GRD} \approx 1.22 \frac{\lambda R}{D}
\end{equation}

For a digital imaging device, the ground sample distance ($\mathit{GSD}$) describes the smallest distinguishable element on an acquired optical image resulting from the pixel sampling on the image plane. This can be calculated by considering the range to the target $R$, the pixel size $x$, and the focal length $f$ of the telescope.
\begin{equation}
\mathit{GSD} = \frac{xR}{f}
\end{equation}

As the orbital altitude is reduced the following relationships regarding the spatial resolution can therefore be identified:
\begin{enumerate}[label=\roman*.]
\item For a fixed aperture diameter $D$, spatial resolution (both $\mathit{GRD}$ and $\mathit{GSD}$) is improved by reducing altitude.
\item The aperture diameter $D$ can be made smaller with reducing altitude whilst maintaining a fixed spatial resolution.
\end{enumerate}

\subsubsection{Modulation Transfer Function} \label{S:MTF}
Further constraints on the spatial resolution arise from the geometry of the sensor and detecting elements, and any optical or aberrations not accounted for in the system design. However, these parameters are generally not linked to change of orbital altitude. 

A parameter known as the \emph{modulation transfer function} (MTF) can be used to characterise other conditions which affect the image quality. The MTF describes the sensitivity of the imaging chain to the spatial frequency or distribution of imaged objects and can be expressed as the variation in contrast or modulation depth (difference between maximum and minimum amplitude) of a sinusoidal image pattern between the object plane and the focal plane \cite{Schott2007}. 

As a result of the blurring and convolution of the PSFs of neighbouring points, the effects of limited spatial resolution become more significant at high spatial frequencies, ie. objects which are spaced closer together become harder to distinguish from each other. The MTF is therefore often seen as a decreasing function of spatial frequency and can be described as the ability of an optical system to capture or transfer contrast in an image at a given resolution. 

The total MTF of a system can be determined by multiplication (cascading) of the individual contributing MTFs which each have a value in the range in value from 0 to 1. For an orbiting spacecraft platform the total MTF is principally comprised of the components relating to the optics, detector, motion, atmosphere, and platform stability \cite{Schott2007}. 

The combined MTF of the system is largely independent of the altitude at which the platform is operated. However, a reduction in altitude can have some impact on components of the MTF:
\begin{enumerate}[label=\roman*.]
\item Assuming a fixed optical aperture and detector size, a reduction in altitude will negatively impact the motion MTF contribution as the ground-speed of the spacecraft increases \cite{Schott2007}. However, as the radiometric performance is also improved by a reduction in altitude (discussed in \cref{S:RadiometricPerformance}), the integration or exposure time can also be reduced, resulting in either a similar or improved motion MTF contribution.
\item For a fixed angular field of regard, the atmospheric MTF performance will not vary with altitude. However, if the total coverage area is to be maintained, a reduction in altitude will increase the path distance through the atmosphere and a worse atmospheric MTF performance at the edge of the field of regard will be experienced in comparison to higher orbits \cite{Schott2007,Sadot1995,Sadot1993}.
\item Lower altitude orbits may be associated with additional periodic and non-periodic disturbances related to the atmospheric density and may therefore experience a degradation in MTF performance related to the vibrational response of the platform \cite{Wulich1987,Haghshenas2017}. However, further work is required in this area to characterise the small-scale variations in atmospheric density and the potential effect on spacecraft structural dynamics.
\end{enumerate}

\subsection{Radiometric Performance} \label{S:RadiometricPerformance}
The radiometric resolution of a system describes the depth of information which is captured in an image.  The radiometric depth is typically measured in number of bits, representing in the number of different brightness, intensity, or colour levels which can be resolved by the sensor. This radiometric performance is principally dependent on the amount of signal which is received at the detector and the sensitivity of the equipment to the magnitude of the electromagnetic energy to which it is exposed. The ratio between the largest and smallest of these levels is correspondingly known as the dynamic range of the sensor.

In general, the power or intensity of an emitted signal received at a given distance $R$ (in a vacuum) follows the inverse-square law as it evenly radiated from a point into three-dimensional space.
\begin{equation}
P \propto \frac{1}{R^2}
\end{equation}

The power of a signal received in orbit is therefore proportional to inverse square of the distance of the spacecraft to the target, which is related to the orbital altitude. For active sensor types where a signal is both transmitted and subsequently received by the spacecraft (eg. radar) this relationship becomes proportional to the fourth power of the range to the target. A reduction in orbit altitude can therefore significantly increase the power received at the sensor and may allow less sensitive detectors or antennae to be used whilst maintaining similar performance. Alternatively, the exposure or dwell time of the sensor may be shortened whilst maintaining the radiometric performance.

The power received by a sensor is also proportional to the collection area. For a circular aperture, for example a telescope, the power is proportional to the square of the diameter $D$. 
\begin{equation}
P \propto D^2
\end{equation}

As the orbital altitude is reduced the collection aperture can therefore be reduced whilst maintaining a similar radiometric performance.

The signal-to-noise (SNR) ratio received at the sensor or detector is often used to characterise the radiometric performance and can be determined by considering the ratio of signal power received from the source $S(\lambda)$ and the different components of noise $N_x(\lambda)$ as a function of wavelength \cite{Silny2014}.
\begin{equation}
\mathit{SNR}_{\mathit{total}}(\lambda) = \frac{S_{\mathit{target}}(\lambda)}{\sum N_x(\lambda)}
\end{equation}
or in units of decibels and generalised to total power:
\begin{equation}
\mathit{SNR}_{\mathit{dB}} = 10 \log_{10} \left( \frac{P_{\mathit{signal}}}{P_{\mathit{noise}}} \right)
\end{equation}

A SNR of greater than 1 is typically necessary to ensure that any signal can be discerned from the background noise levels. However, for precise and accurate measurements or high-quality imagery greater SNRs are often required.

Alternatively, different noise-equivalent metrics can also be used, for example:

\begin{enumerate}[label=\roman*.]
	\item The noise-equivalent delta in reflectance ($\mathit{NE\Delta\rho}$), ie. the smallest difference in surface reflectance that changes the signal by a value equal to the magnitude of the total noise.
	\item The noise-equivalent delta in emittance ($\mathit{NE\Delta\varepsilon}$), ie. the smallest difference in surface emittance that changes the signal by a value equal to the magnitude of the total noise.
\end{enumerate}

Sources of noise include those which are related to the sensor and associated electronics and additional components which arise from the contrast between the background and the target and the atmospheric path through which the signal passes. As with the atmospheric contributions MTF, the effect on SNR is principally due to the scattering and adsorption effects of water vapour and other aerosols in the lower strata, leading to signal attenuation. 

The SNR will therefore be affected by both the total range to the target and the atmospheric path length. The SNR is therefore related to the orbiting altitude and off-nadir angles used by the imaging system. Radiometric performance is positively affected by a reduction in orbital altitude as the power/signal received improves with reduced range to the target. However, for significant off-nadir viewing angles the SNR performance may degrade as the range to target through the lower atmosphere increases.

These relationships are also consistent with the trends identified for diffraction limited resolution. As the orbital altitude is reduced the aperture diameter can be made smaller whilst broadly maintaining the same radiometric performance and spatial resolution. Alternatively, if the same aperture diameter is maintained the radiometric performance and spatial resolution will be improved with reducing altitude.

\subsection{Temporal Resolution}
Revisit time, defined as the time for a satellite to acquire successive viewings of target locations on the Earth’s surface can highly influence satellite constellation design, configuration, performance and technology selection \cite{Smith2007}. Low maximum revisit time (MRT) is generally desired along with global coverage (possibly between prescribed latitude bands) for most Earth observation and remote sensing applications.

The selection of a repeating ground-track orbit is not strictly required for EO missions, but it represents a critical requirement when regular passes over specific locations are fundamental for achieving the mission objectives. Short-period repeating ground-tracks are more easily achievable for higher LEO altitudes (\SIrange{600}{1000}{\kilo\meter}), where external perturbations typically have less effect on the orbital dynamics. In the VLEO altitude range, the residual atmospheric interaction with the satellite’s external surfaces gradually causes the orbit to decay, thus making repeating ground-tracks only achievable when aerodynamic compensation devices or aerodynamic control techniques are employed \cite{VirgiliLlop2015}.

\begin{figure}
	\centering
	\begin{subfigure}[b]{\linewidth}
	\centering
	\includegraphics[width=90mm]{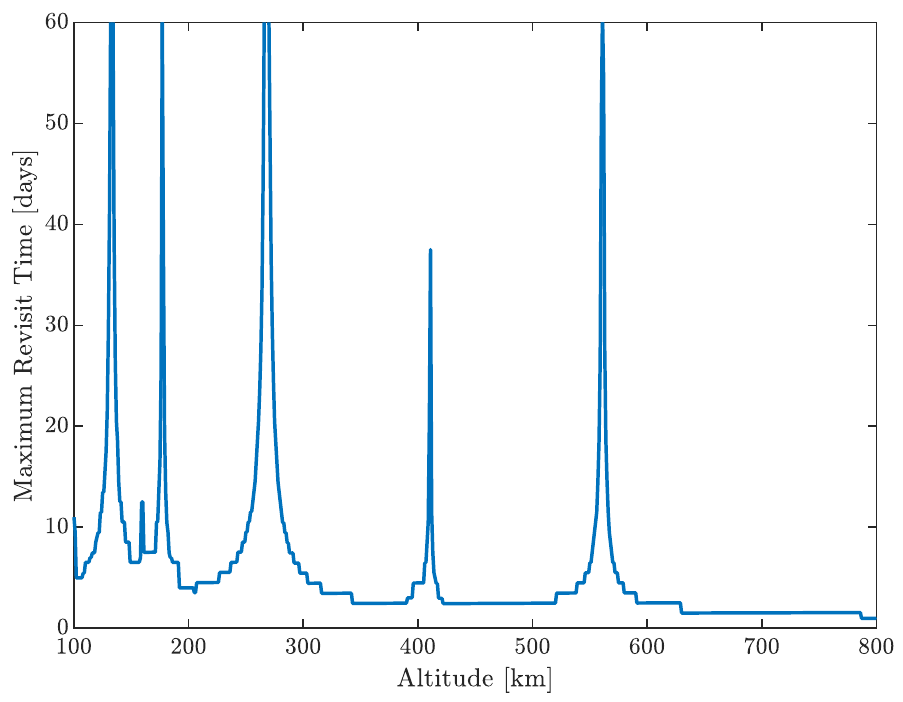}
	\caption{Sun-Synchronous Orbit.}
	\label{F:MRT_SSO}
	\end{subfigure}
	\\
	\begin{subfigure}[b]{\linewidth}
	\centering
	\includegraphics[width=90mm]{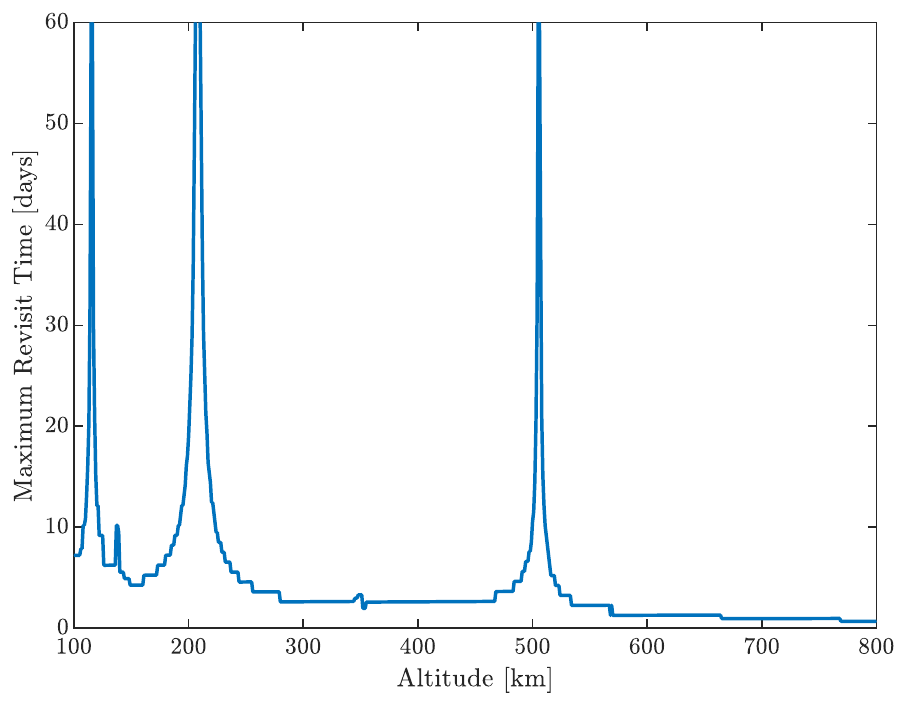}
	\caption{Non-SSO ($i=$\ang{60}).}
	\label{F:MRT_NSSO}
	\end{subfigure}
	\\
	\begin{subfigure}[b]{\linewidth}
	\centering
	\includegraphics[width=90mm]{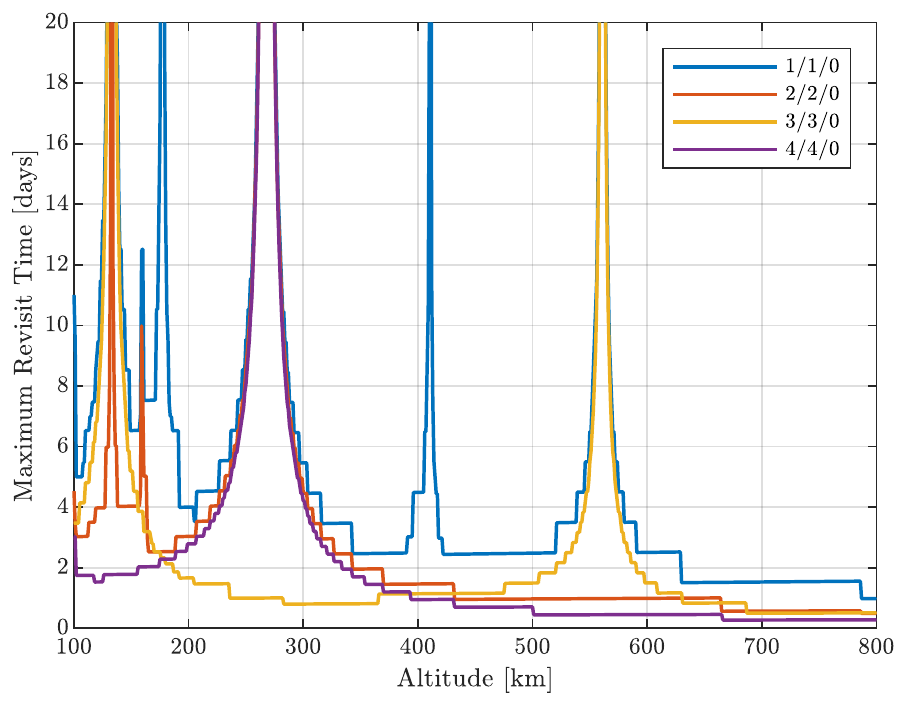}
	\caption{Varying constellation configuration in SSO.}
	\label{F:MRT_Constellation}
	\end{subfigure}
	\caption{Maximum Revisit Time (MRT) at \ang{40} target latitude for varying altitude ($\psi=\ang{45}$). Calculated using the semi-analytical method of \citet{Crisp2018}.}
	\end{figure}

\cref{F:MRT_SSO} shows how MRT varies according to altitude for circular sun-synchronous orbits (SSOs), when the field of regard angle $\psi$ is equal to \ang{45} and the target latitude is \ang{40}. Some altitude ranges provide poor temporal resolution, for example about approximately \SI{130}{\kilo\meter} and \SI{175}{\kilo\meter} as indicated by the peaks in \cref{F:MRT_SSO}, and should be avoided where regular and complete coverage of a given latitude range is required. The temporal performance of these altitudes typically results from resonance of the orbital period with the rotation period of the Earth leading to incomplete coverage of all latitudes, or very long repeat ground-track patterns.

For the VLEO range, close to optimal MRT is still achievable for certain altitude windows, as shown in \cref{F:MRT_SSO}, suggesting that there are restrictions on the usable altitude ranges which the satellite can operate effectively in. Currently, the majority of the EO missions are launched into SSO for the advantageous illumination conditions these orbits offer. However, it is worth mentioning that the non-SSOs can provide improved temporal resolution in the VLEO range, for example as indicated in \cref{F:MRT_NSSO} where the altitude range from approximately \SIrange{250}{475}{\kilo\meter} is shown to provide low MRT. 

Temporal resolution can introduce some constraints on the altitude windows in which LEO satellite constellations can be operated. The low MRT achievable for the range (\SIrange{600}{800}{\kilo\meter}) generally makes these altitude windows suitable for most EO missions. For certain ranges, small changes in altitude can result in significant variation in temporal resolution performance. However, it is interesting to notice how SSO constellations consisting of an odd number of planes, each occupied by a single satellite, can provide significant improvement for certain lower altitude windows, granting comparable performance in terms of temporal resolution to higher altitudes. In \cref{F:MRT_Constellation} this is demonstrated by the low MRT for a Walker Delta configuration\footnotemark of 3/3/0 over the altitude range of (\SIrange{200}{350}{\kilo\meter}).

\footnotetext{Typically described as \textbf{i:t/p/f} where \textbf{i} is the inclination, \textbf{t} the total number of satellites, \textbf{p} the number of planes, and \textbf{f} the relative spacing between satellites in adjacent planes.}

\subsection{Ground Communication and Link Budget}
The communications performance of a space system is dependent on the location of available ground stations and in-orbit networks, the orbital parameters, and the subsystem and hardware selection. Variation in the orbit altitude can therefore have an impact on the overall communications performance which can be achieved in orbit.

The radiometric performance of data communications (receiving and transmitting data) is broadly similar to the relationships described in \cref{S:RadiometricPerformance}. Principally, the free-space loss reduces with the shorter range to the target ground-station (inverse-square law \cite{Wertz2011}) and the signal-to-noise ratio therefore improves for an antenna of the same size and effective isentropic radiated power (EIRP). The radiated power $P_r$ received at the range R is dependent on the receiving antenna diameter $D$ and the antenna efficiency $\eta_{\mathit{ant}}$.
\begin{equation}
P_r = \mathit{EIRP} \cdot \frac{L_a D^2 \eta_{\mathit{ant}}}{16 R^2}
\end{equation}

The transmission path loss factor $L_a$ includes absorption due to the ionosphere, atmosphere, and rain, and is therefore dependent on the slant range through the atmosphere or the off-nadir angle. Thus, for lower altitude orbits, the allowable elevation angle to ensure a reliable communication link may become a limiting factor and could reduce the effective access time \cite{Ippolito1986}. Further sources of noise should also be considered in the calculation of SNR, including those originating in the antenna and external sources of electromagnetic radiation including solar, cosmic background, and from the Earth (both natural and man-made) \cite{Fortescue2011}.

In addition, the velocity increase with reducing orbital altitude may also adversely affect the available duration for communication with a given ground station, reducing the volume of data which can be transferred.

Finally, the frequency of passes within range of the available ground stations should be considered for different orbital altitudes as per the discussion in Section 2.1.4. If fewer passes of available ground stations are performed each day for a lower altitude orbit, the total volume of data which can be transmitted or received may be constrained despite the improvement in radiometric performance.

\subsection{Deorbit Requirements}
The Inter-Agency Space Debris Coordination Committee (IADC) guidelines define LEO (up to \SI{2000}{\kilo\meter}) as a protected region of Earth orbit and outline that any spacecraft operating in this region should either be deorbited after the completion of operations or have a maximum lifetime of less than 25 years \cite{Inter-AcencySpaceDebrisCoordinationCommittee:SteeringGroupandWorkingGroup42007}. These recommendations are also incorporated into space agency requirements and policy (eg. NASA \cite{NASAOfficeofSafetyandMissionAssurance2017,NASAOfficeofSafetyandMissionAssurance2019}, ESA \cite{EuropeanSpaceAgencyDirectorGeneralsOffice2014}) and an ISO Standard (24113:2011 Space systems – Space debris mission requirements) \cite{ISOTC2010}. However, whilst these guidelines and associated standard are not technically law or regulation, they are often used by national agencies and governing authorities when considering whether to grant launch licenses and must therefore be satisfied in most cases to successfully gain access-to-orbit.

The post-mission orbital lifetime of a spacecraft in LEO can extend to many hundreds or thousands of years depending on the orbital parameters and physical properties of the satellite. However, as orbital altitude is reduced the residual atmosphere of the Earth becomes denser the lifetime quickly decreases. Lower Earth orbits are therefore more likely to directly comply with the deorbit requirements.

Estimation of orbital lifetime is a difficult process due to the uncertain nature of the thermospheric density. Whilst a number of atmospheric density models are openly available, these all have uncertainties, bias, and errors which can significantly affect the calculated orbital lifetime. Many of these models are also highly dependent on the predicted solar cycle which is a principal driver of the variation in atmospheric density. Forecasts or models for future solar activity are also available, but often it is found that long-term predictions are inaccurate and unable predict the correct trend in solar activity \cite{Vallado2008}. Orbital lifetime prediction methods can also vary significantly in fidelity depending on the type of propagation performed (ie. analytical or numerical), scope of the perturbations included, and the input data used.

\begin{figure}
	\centering
	\includegraphics[width=90mm]{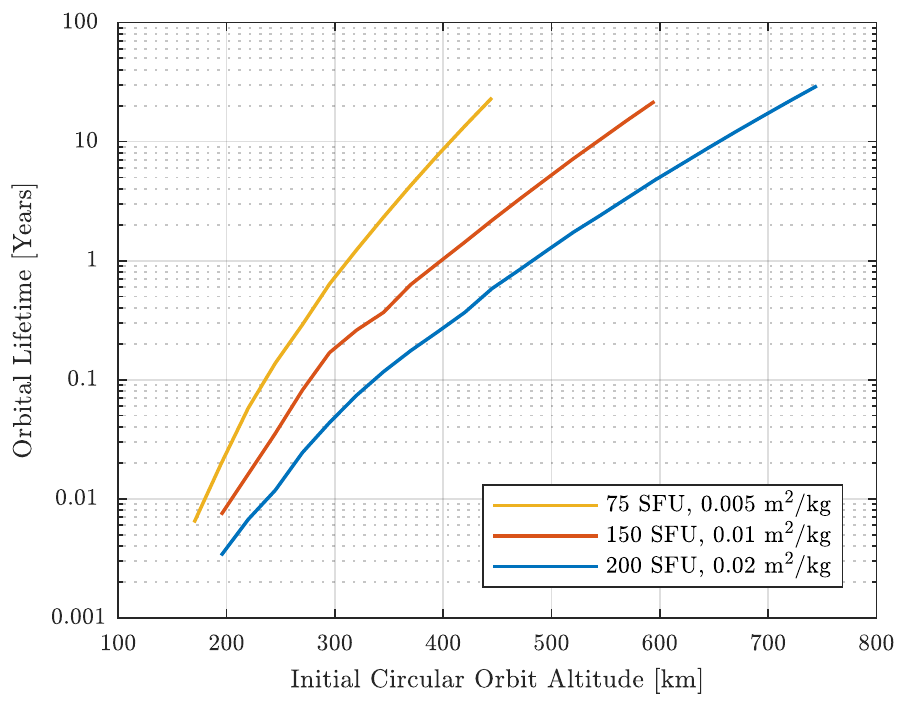}
	\caption{Range of orbital lifetime for different initial circular altitudes dependent on combination of ballistic coefficient and solar radio flux conditions. Generated using a semi-analytical propagation method (SALT \cite{Liu1980}) and the NRLMSISE-00 \cite{Picone2002} atmosphere model.}
	\label{F:OrbitLifetime}
	\end{figure}

However, by using simple analytical or semi-analytical methods the expected range of orbital lifetime for spacecraft in LEO for different altitudes can be illustrated. \cref{F:OrbitLifetime} shows this significant variation in orbital lifetime which occurs with altitude, solar flux input to the atmosphere, and physical satellite characteristics (ballistic coefficient or mass to area ratio assuming a typical drag coefficient of \num{2.2}). For example, at an altitude of \SI{400}{\kilo\meter} the lifetime can be shown to vary from approximately \SI{3}{months} for a spacecraft with a low ballistic coefficient and under high solar flux conditions to over \SI{8}{years} for a characteristically high ballistic coefficient spacecraft and under low solar flux conditions. If significantly lower coefficients of drag can be produced, for example through development of low drag materials and geometries \cite{Roberts2017}, the lifetime for all altitudes subject to atmospheric drag in LEO can be increased.

The results presented broadly demonstrate that any satellite operating in a VLEO orbit (below \SI{450}{\kilo\meter}) will have a post-mission lifetime of less than 25 years, regardless of the solar environment and satellite size and mass. \citet{Oltrogge2007} present a different approach for lifetime analysis which utilises a random draw method for the solar flux, but similarly show that orbits below \SI{500}{\kilo\meter} generally have a lifetime of less than 25 years.

VLEO orbits are therefore generally compliant with the IADC guidelines and corresponding licensing requirements. Furthermore, this compliance is not conditional on any additional deorbit hardware or propulsion system which can add complexity, cost, and system mass.

\subsection{Debris Collision Risk Resilience}
The debris environment which exists in Earth orbit is predominantly a result of the exploration and operational activities which have occurred since the beginning of human involvement in space. In addition to naturally occurring micrometeoroids the objects which persist in orbit principally include post-mission and failed spacecraft, launch vehicle upper stages, deployment and other mission-related items, and surface degradation and propulsion products amongst other miscellaneous objects \cite{Kilinkrad2006}. Explosions, collisions, and break-up or fragmentation events within this population have further increased the number and dispersion of these objects in the orbital environment.

In LEO, the residual atmospheric environment causes these objects to decay, eventually causing re-entry. However, as the atmospheric density reduces roughly exponentially with altitude, the rate of decay from upper and mid LEO is slow, and the lifetime of debris can often exceed the mission lifetime of many spacecraft. However, in VLEO the atmospheric density is higher and any debris which is generated in or enters this regime from higher orbits will decay at a faster rate.

Prediction of the future space debris environment can be generated using ESA’s MASTER-2009 (Meteoroid and Space Debris Terrestrial Environment Reference) tool \cite{Flegel2011}. This tool provides the capability to predict the spatial density or flux of known debris sources (greater than \SI{1}{\micro\meter}) against a target spacecraft surface/volume. The future debris population can be modelled assuming either a ``business-as-usual'' case or with under the different debris mitigation scenarios (eg. including explosion prevention and spacecraft end-of-life disposal).

\begin{figure} 
	\centering
	\includegraphics[width=90mm]{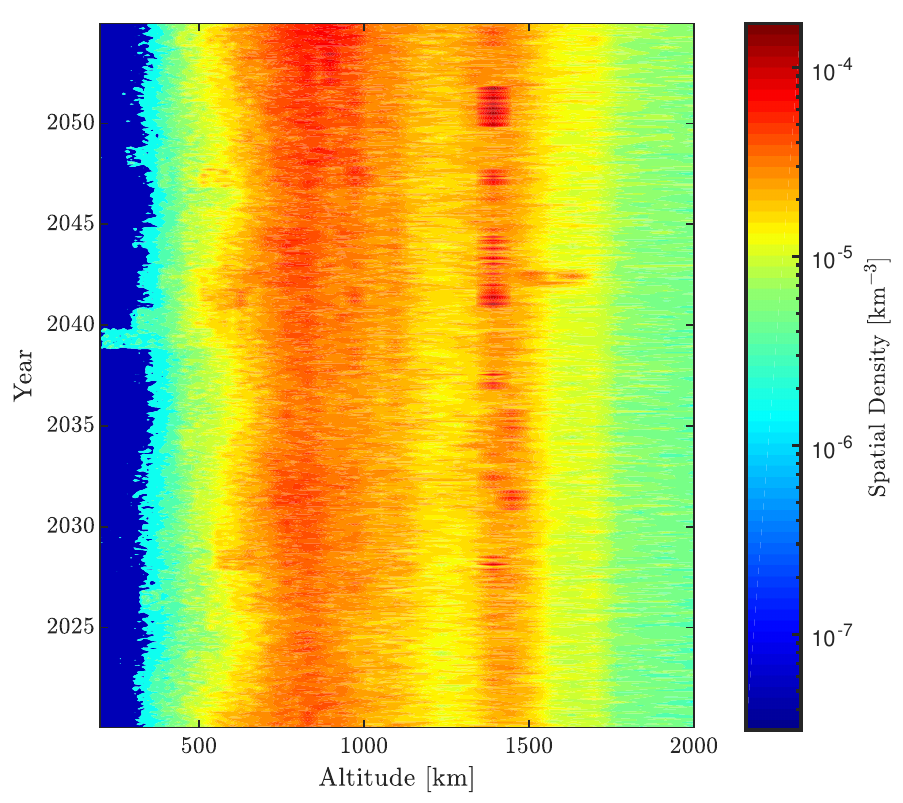}
	\caption{Simulated LEO debris population (spatial density) over the period 2020 to 2055 based on a “Business-as-usual” scenario with no debris mitigation. Generated using the ESA MASTER-2009 tool \cite{Flegel2011}.}
	\label{F:DebrisPop}
	\end{figure}

\begin{figure}
	\centering
	\includegraphics[width=90mm]{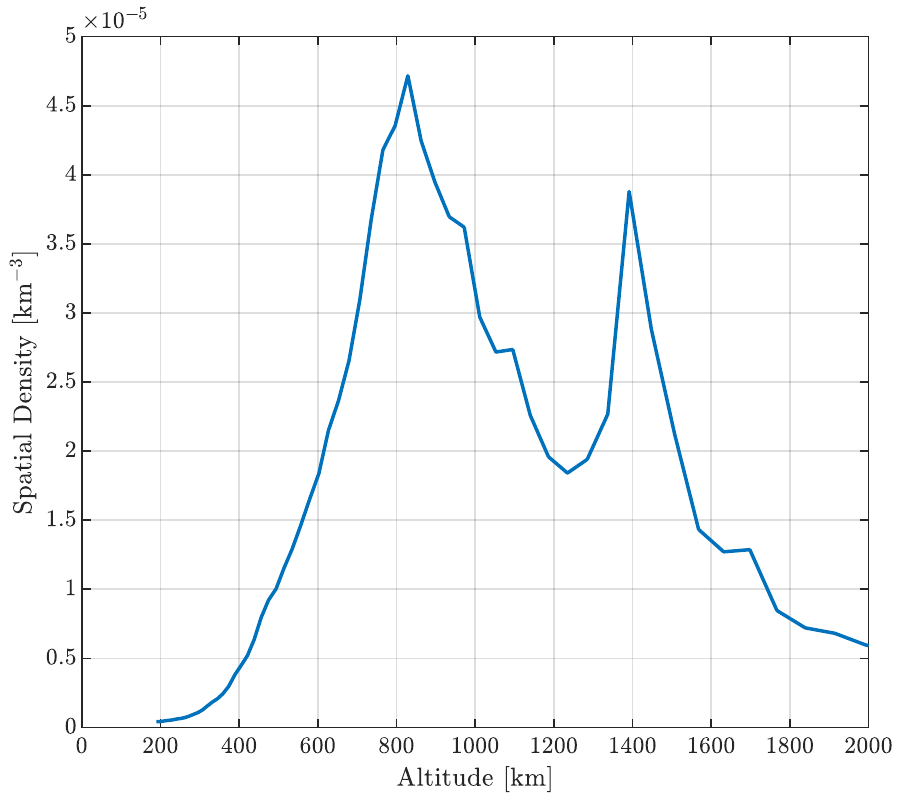}
	\caption{Spatial density for LEO altitudes averaged over the period 2020 to 2055.}
	\label{F:DebrisDensity}
	\end{figure}

The projected debris population at different altitudes for a ``business-as-usual'' case is shown in \cref{F:DebrisPop} over the period 2020 to 2055. The average spatial density over this period is calculated for each altitude and shown in \cref{F:DebrisDensity}. In both cases, the VLEO range ($<$ \SI{500}{\kilo\meter}) is clearly shown to have a lower spatial density profile than higher LEO orbits and appears to be resilient to the debris build up which is predicted for the \SIrange{700}{1000}{\kilo\meter} range towards 2055.

It should be noted that the modelled future populations in MASTER-2009 do not include the recent mega-constellations that have been planned or are currently being launched (eg. OneWeb, SpaceX Starlink, Amazon ``Project Kuiper''). However, as these systems are planned for orbital altitudes above that of VLEO, the risk of collision in the VLEO altitude range remains low and is only affected by those spacecraft that need to de-orbited or that have suffered complete failure and will naturally decay through the VLEO range.

The assessment of collision risk in orbit includes both the probability and potential consequence and is a combination of the spacecraft composition and geometry, the debris environment, and the relative collision velocity \cite{ScientificandTechnicalSubcommitteeoftheUnitedNationsCommitteeonthePeacefulusesofOuterSpace1999}.  Using the spatial density profiles presented in \cref{F:DebrisPop,F:DebrisDensity} to describe the probability of a collision at a given altitude, the VLEO range is shown to remain at a lower relative risk than other orbits which may become over-populated by the aforementioned mega-constellations. Similarly, the risk to VLEO in the case of a Kessler syndrome type cascade event will remain relatively low as any debris that is generated in VLEO or enters from a higher altitude will quickly decay through the range and deorbit. However, the risk profile to spacecraft operating in VLEO will still increase to a small extent in this scenario due to the increased flux of de-orbiting spacecraft and debris which transits through the VLEO range.

\subsection{Radiation Environment}
The radiation environment which a spacecraft is subjected to in LEO consists of a combination of energetic particles trapped by the Earth’s magnetic field, solar flares, and galactic cosmic rays \cite{Stassinopoulos1988}. These sources of radiation can interact with the sensitive components of spacecraft subsystems causing both long-term and single-event effects which can have significant detrimental effect on a spacecraft mission. Radiation-hardened electronic components or fault-resilient software systems are therefore typically employed at significant additional cost. Radiation-shielding can also be employed to reduce the radiation dosage which internal components are exposed to \cite{Wertz2011}.

The radiation environment can also affect the performance and longevity of materials (eg. polymer embrittlement) used on a spacecraft \cite{Finckenor2015}. For long-duration missions, alternative material choices or design redundancy may therefore be required to ensure structural integrity, possibly increasing system mass and cost.

The radiation environment in Earth orbit is characterised by the presence of the magnetosphere and the Van Allen radiation belts \cite{Fortescue2011}. The exposure due to the trapped-radiation in the Van Allen belts is known to vary broadly with the solar cycle and can be modelled by NASA’s AP-8 and AE-8 models for proton and election content respectively at either the maximum or minimum solar conditions \cite{Sawyer1976,Vette1991}.

\begin{figure*} 
	\centering
	\includegraphics[width=\textwidth,trim={30 10 30 10},clip]{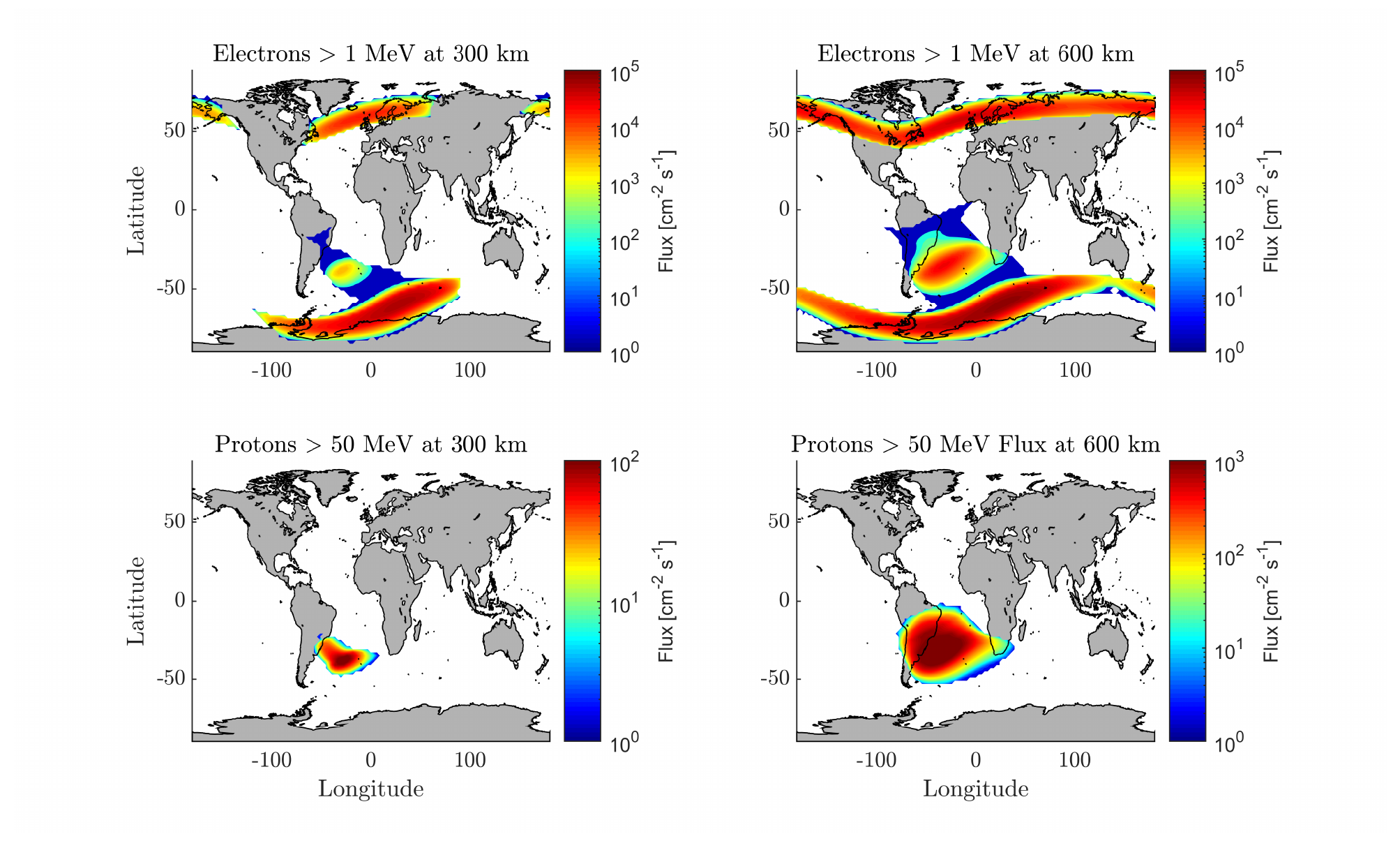}
	\caption{Proton (AP-8 model) and electron (AE-8 model) flux at \SI{300}{\kilo\meter} and \SI{600}{\kilo\meter} altitude at solar maximum conditions. Data generated using ESA SPENVIS online tool \cite{Heynderickx2000}.}
	\label{F:RadMaps}
	\end{figure*}

In the LEO range, the distribution of protons and electrons of varying energy level can be calculated using these models through an online tool, ESA's Space Environment Information System (SPENVIS) \cite{Heynderickx2000}, and represented globaly in \cref{F:RadMaps}. For the flux of electrons, the peak magnitude is not found to decrease with a reduction in altitude, but the geographic distribution of electrons with high energy can be seen to decrease. Correspondingly, the peak proton flux is shown to reduce by an order of magnitude between \SI{600}{\kilo\meter} and \SI{300}{\kilo\meter} whilst the distribution of substantial flux is also shown to reduce significantly. This is in part due to the increasing atmospheric density \cite{Badhwar1997,Badhwar1999}, and thus demonstrates a benefit in operating at a lower orbital altitude.

With increasing interest and use of commercial off-the-shelf components \cite{Boshuizen2014,Tyc2005,Underwood2001,Sweeting2018} without radiation-hardening, a reduction in radiation exposure at lower altitudes may enable longer duration missions utilising these components as the lifetime dosage reduces correspondingly. Alternatively, even cheaper consumer components may be able to be successfully used in VLEO, further decreasing mission costs and system development time.

\subsection{Access to Orbit}
\begin{figure}
	\centering
	\includegraphics[width=90mm]{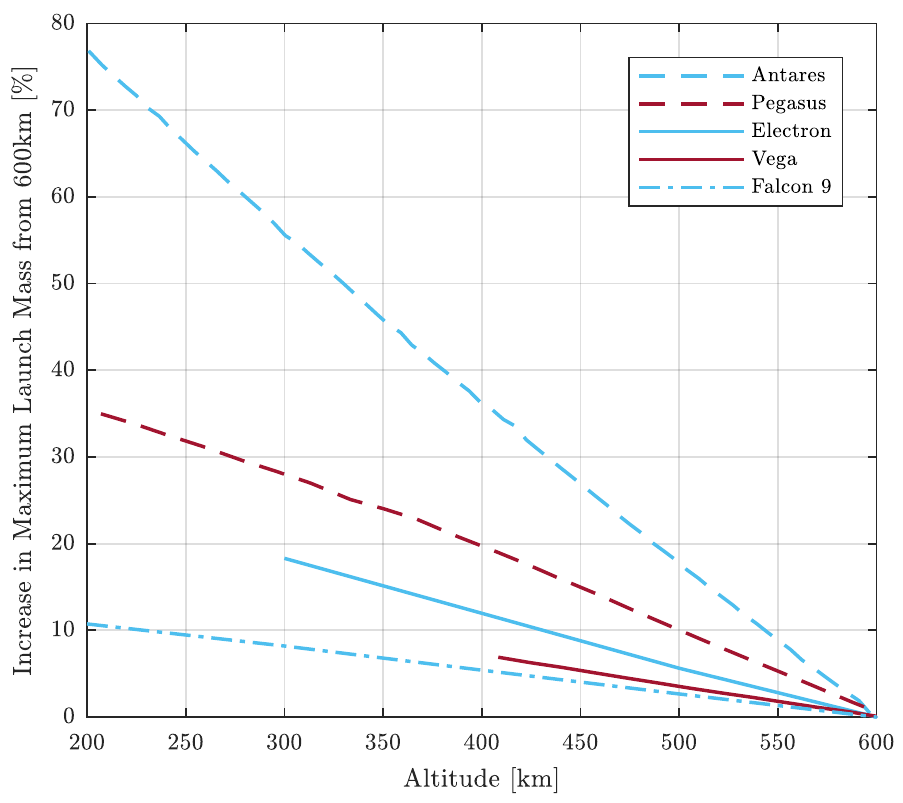}
	\caption{Variation in SSO launch capability with decreasing altitude for different launch vehicles \cite{SpaceExplorationTechnologies2015,OrbitalSciencesCorporation2013,RocketLabUSA2016,OrbitalSciencesCorporation2015,Arianespace2014}.}
	\label{F:LaunchMass}
	\end{figure}

The payload performance or launch mass of an orbital launch vehicle generally increases with reducing altitude, principally due to lower gravity losses and shorter flight-times and therefore reduced fuel requirements. A comparison of the SSO launch capability of different vehicles (Falcon 9 \cite{SpaceExplorationTechnologies2015}, Antares \cite{OrbitalSciencesCorporation2013}, Electron \cite{RocketLabUSA2016}, Pegasus \cite{OrbitalSciencesCorporation2015}, and Vega \cite{Arianespace2014}) is shown in \cref{F:LaunchMass}, illustrating the increase in launch performance which can be achieved for lower altitude insertion orbits.

For different vehicles, the improvement in launch capability from \SI{600}{\kilo\meter} and \SI{300}{\kilo\meter} insertion ranges from approximately 10\% to over 50\%, demonstrating potential for a significant increase in mass which can be launched to lower altitude orbits.

A greater number of satellites can therefore be delivered per launch to orbit for no additional cost. Alternatively, for shared launch services, the unit cost (cost per \si{kg} or per satellite of a given mass) can be decreased, improving the accessibility of VLEO missions.

Finally, as each vehicle has a greater payload capability to lower altitude orbits the number of vehicles which can launch a given spacecraft may be increased, thus increasing competition and providing alternative options in the case of potential launch delays.

\subsection{Geospatial Position Accuracy}
The error between the reported/recorded and actual location of an acquired image or other Earth observation measurement is generally referred to as the geospatial or geometric position accuracy. The principal contributors to errors in geospatial position are the uncertainty in the spacecraft position and attitude, and errors associated with the alignment and calibration of these sensors and any observing instruments. A distributed set of ground control points are often used to provide correction to acquired data and imagery, improving the geospatial position accuracy.

The principal geospatial errors are associated with the satellite position (in-track, cross-track, and radial), and pointing (elevation/nadir and azimuthal). Additional errors also arise from the uncertainty in the altitude of the observed target, and uncertainty in the rotational position of the Earth due to clock errors. 

Mapping errors are described by \cref{E:MappingErr} \cite{Wertz2011}, where $h_T$ is the altitude of the target on the Earth’s surface, $\phi$ is the target latitude, and $\varphi$ is the azimuth of the target with respect to the satellite ground-track . Referring back to the geometry in \cref{F:FootprintGeometry}, $R$ is the slant range to the target, $\psi$ is the angle of the target from the spacecraft nadir, $\varepsilon$ is the elevation angle of the spacecraft from the target, and $\theta$ is the Earth central angle. The magnitude of each of the respective errors is indicated by the various $\Delta_{x}$ parameters.
\begin{subequations} \label{E:MappingErr}
\begin{align}
E_{\mathit{m,azimuth}} &= \Delta_\varphi \cdot R \sin{\psi} \\
E_{\mathit{m,elevation}} &= \Delta_\psi \cdot\frac{R}{\sin{\varepsilon}} \\
E_{\mathit{m,in-track}} &= \Delta_I \cdot \frac{R_{\phi} + h_T}{r_s} \sqrt{1-(\sin{\theta \sin{\varphi}})^2} \\
E_{\mathit{m,cross-track}} &= \Delta_C \cdot \frac{R_{\phi} + h_T}{r_s} \sqrt{1-(\sin{\theta \cos{\varphi}})^2} \\
E_{\mathit{m,radial}} &= \Delta_{r_s} \cdot \frac{\sin{\psi}}{\sin{\varepsilon}} \\
E_{\mathit{m,altitude}} &= \Delta_{(R_{\phi} + h_T)} \cdot \frac{1}{\tan{\varepsilon}} \\
E_{\mathit{m,clock}} &= \Delta_t \cdot V_E \cos{(\phi)}
\end{align}
\end{subequations}
The mapping errors due to satellite position error (in-track, cross-track) show a weak proportional relationship to the orbital altitude, whilst the corresponding errors associated with the satellite attitude (azimuth and elevation) demonstrate a stronger proportional relationship with the range to the target. Errors associated with radial position error and uncertainty in the altitude of the target are trigonometric functions of the elevation angle and therefore also related to the orbit altitude for a given off-nadir angle. Errors associated with the on-board clock or timing do not demonstrate any dependence on altitude.

\begin{figure}
	\centering
	\includegraphics[width=90mm]{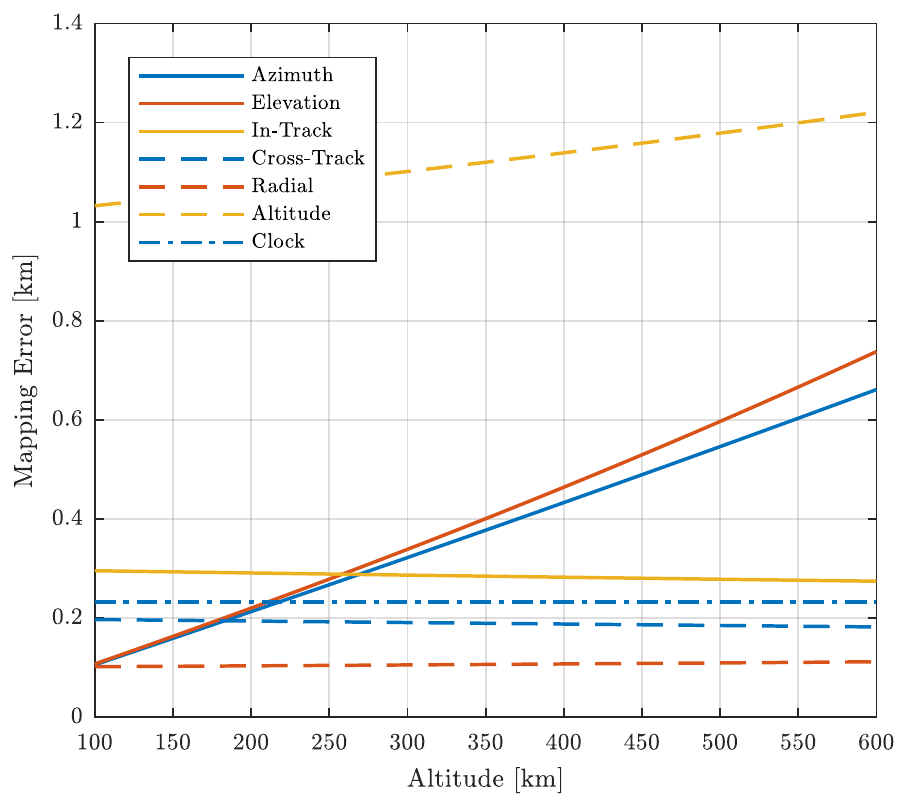}
	\caption{Mapping error for an equatorial ground target with varying altitude at an off-nadir pointing angle of \ang{45} (Error sources: $\Delta\phi=\ang{0.06}$, $\Delta\eta=\ang{0.03}$, $\Delta I=\SI{0.3}{\kilo\meter}$, $\Delta C=\SI{0.2}{\kilo\meter}$, $\Delta R_T=\SI{0.1}{\kilo\meter}$, $\Delta T=\SI{0.5}{\second})$.}
	\label{F:MappingErr}
	\end{figure}

The relationship between these different error sources and the resulting geospatial position accuracy with reducing orbital altitude is illustrated in \cref{F:MappingErr} using representative values for the individual error sources. The trends in errors associated with the pointing/attitude errors demonstrate that the geospatial position accuracy generally improves with reducing orbital altitude. Alternatively, given the trend between the attitude errors (azimuth and elevation) with the mapping error, the pointing requirements for a platform can be relaxed with a reduction in altitude.

\begin{figure}
	\centering
	\includegraphics[width=90mm]{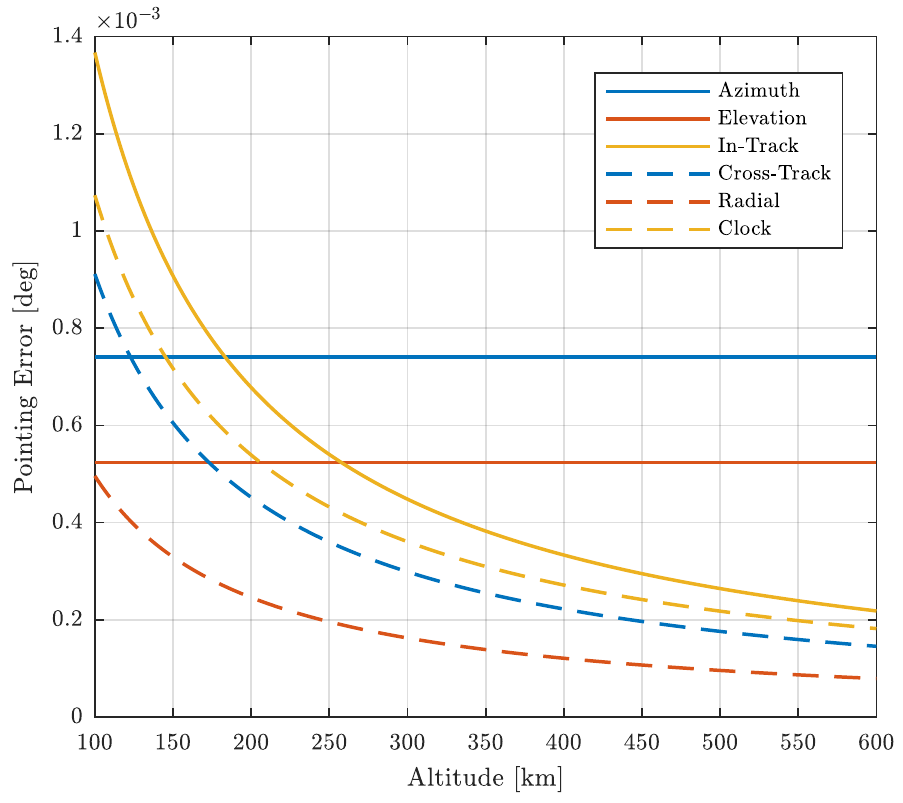}
	\caption{Pointing error for an equatorial ground target with varying altitude at an off-nadir pointing angle of \ang{45} (Error sources: $\Delta\phi=\ang{0.06}$, $\Delta\eta=\ang{0.03}$, $\Delta I=\SI{0.3}{\kilo\meter}$, $\Delta C=\SI{0.2}{\kilo\meter}$, $\Delta R_T=\SI{0.1}{\kilo\meter}$, $\Delta T=\SI{0.5}{\second})$.}
	\label{F:PointingErr}
	\end{figure}

Error in the pointing of a spacecraft towards a given target are are similarly described by \cref{E:PointingErr} \cite{Wertz2011} and indicated in \cref{F:PointingErr}. 
\begin{subequations} \label{E:PointingErr}
\begin{align}
E_{\mathit{p,azimuth}} &= \Delta_\varphi \cdot \sin{\psi} \\
E_{\mathit{p,elevation}} &= \Delta_\psi \\
E_{\mathit{p,in-track}} &= \Delta_I  \cdot \frac{\sin({\cos{\varphi}\cos{\varphi}})}{R}\\
E_{\mathit{p,cross-track}} &= \Delta_C  \cdot \frac{\sin({\cos{\varphi}\cos{\varphi}})}{R}\\
E_{\mathit{p,radial}} &= \Delta_{r_s}  \cdot \frac{\sin{\psi}}{R}\\
E_{\mathit{p,clock}} &= \Delta_t \cdot \frac{V_E}{R}\cos{(\phi)}\sin{(\cos{\varphi_E}\cos{\varepsilon})}
\end{align}
\end{subequations}
As the pointing error is associated with the attitude determination and control capability (azimuth and elevation errors), there is no dependency on the range to the target or altitude. However, errors in the satellite position (in-track, cross-track, and radial) and the on-board clock and demonstrate an inverse relationship with the pointing error of the spacecraft with reducing range. For off-nadir pointing, the position knowledge requirement therefore increases with a reduction in spacecraft altitude.

In general, a reduction in orbital altitude therefore reduces the requirements on attitude determination and control. However, the requirement for spacecraft position knowledge may increase modestly. The magnitude of the errors associated with the attitude and position of the spacecraft are based on the available sensors and orbit/attitude determination capabilities. Factors which can affect the accuracy of these sensors may also subsequently affect the geospatial position accuracy of acquired imagery and data. For example, evidence of ionospheric interference of GPS devices in low Earth orbits resulting in tracking losses has been observed, particularly at high-latitudes and periods of high solar activity \cite{VandenIJssel2011}.

\subsection{Aerodynamic Control}
In LEO the interaction between the residual gas particles and the external surfaces of a spacecraft results in the generation of aerodynamic forces and torques. The principal force generated is drag, which acts to cause orbital decay and eventually deorbit. However, out-of-plane forces can also be generated and can contribute to orbital manoeuvring. These forces, in combination with the spacecraft geometry, can also be used to generate torques and used to modify the spacecraft stability and provide attitude control.

A range of different attitude and orbit control methods using these aerodynamic forces and torques have been proposed in literature, but few demonstrated to date. Aerodynamic-based formation-keeping, constellation maintenance, and on-orbit rendezvous manoeuvres were first proposed using only the differential drag force between multiple objects \cite{Leonard1989,Palmerini2005,Bevilacqua2008}. However, more recently, methods exploiting differential lift have emerged \cite{Horsley2013,Traub2019a} and methods using differential drag have been demonstrated in-orbit \cite{Gangestad2013,Foster2016}. Use of drag augmentation has been proposed for targeting of atmospheric re-entry location \cite{VirgiliLlop2015a,Omar2017a,Omar2019} and also collision avoidance \cite{Guglielmo2019}, whilst adjustment of orbital inclination using out-of-plane forces \cite{VirgiliLlop2015} have also been studied. Passive aerodynamic stabilisation or aerostability (the pointing of a spacecraft in the direction of the oncoming flow) has been demonstrated in orbit by several missions \cite{Kumar1995,Drinkwater2007,Sarychev2007}, whilst further aerodynamic attitude control concepts including use of external surfaces to perform detumbling \cite{Hao2016}, internal momentum management \cite{Mostaza-Prieto2017}, and pointing manoeuvres \cite{Chen2000,Gargasz2007,Auret2011,VirgiliLlop2014} have also been considered. Centre-of-mass shifting has also been proposed as a method to augment aerodynamic stabilisation \cite{Chesi2017,Virgili-Llop2019}.

The aerodynamic forces (and torques by association) can be described by the following equation where $\rho$ is the atmospheric flow density, $V$ the relative flow velocity, $A_{\mathit{ref}}$ a reference area, and $C_F$ a corresponding set of force coefficients which are determined by the interaction between the flow and the surface \cite{NASA1971}.
\begin{equation}
\vec{F}_a = \frac{1}{2}\rho \vec{V}_{\mathit{rel}}^2 \frac{\vec{V}_{\mathit{rel}}}{|\vec{V}_\mathit{rel}|} A_{\mathit{ref}} {\vec{C}_F}
\end{equation}

\begin{figure}
	\centering
	\includegraphics[width=90mm]{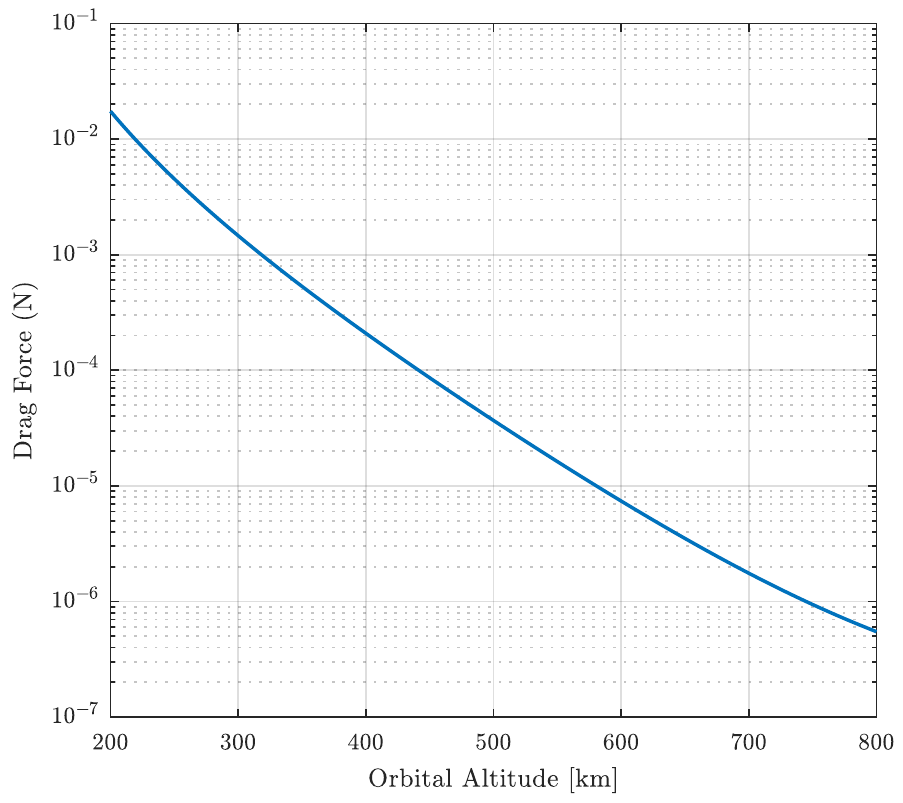}
	\caption{Variation in drag force magnitude with altitude in a circular orbit for a surface oriented normal to the flow of area \SI{1}{\meter\squared} with a coefficient of drag \num{2.2}.}
	\label{F:DragForce}
	\end{figure}

The forces experienced in orbit therefore increase with decreasing orbital altitude as the atmospheric density increases. A further contribution is also provided by the small increase in orbital velocity as altitude decreases. Using the NRLMSISE-00 atmosphere model \cite{Picone2002} with nominal input parameters, the increase in aerodynamic force with altitude, assuming a circular orbit, is shown in \cref{F:DragForce}. The reduction in orbital altitude from \SI{600}{\kilo\meter} to \SI{300}{\kilo\meter} for example is shown to increase the generated force over 200-fold and can therefore result in significantly increased effectiveness or efficiency of aerodynamic attitude and orbit control methods.

Aerostability, for example, has been shown to be possible up to an altitude of approximately \SI{500}{\kilo\meter}, with optimal results demonstrated for altitudes below \SI{450}{\kilo\meter} \cite{Psiaki2004,Armstrong2009,Rawashdeh2013}. This is due to the dependence of aerodynamic stiffness on the residual atmospheric density and the relative magnitude of other perturbing torques, for example due to solar radiation pressure, residual magnetic dipoles, and gravity gradient.

A key consideration in the use of these forces is the ratio between lift (or out-of-plane force) and the drag force. As a result of the rarefied flow environment in LEO and the diffuse gas-surface interactions for typical spacecraft surface materials, this lift-to-drag ratio is generally very low, on the order of \num{0.1} \cite{King-Hele1987,VirgiliLlop2015}. To utilise aerodynamic forces for control and manoeuvring purposes whilst also maintaining a reasonable orbital lifetime this ratio must be increased \cite{Traub2019}, for example through the identification of new materials which promote specular gas reflections, currently an active area of research \cite{Roberts2017}. Alternatively, or additionally, a propulsion system can be utilised to mitigate or counteract the effect of drag, thereby providing an effective increase in lift-to-drag ratio.

\subsection{Atmosphere-Breathing Electric Propulsion}
The increased atmospheric density with decreasing orbital altitude also provides the opportunity to explore atmosphere-breathing propulsion systems. Whilst electric propulsion systems are used widely for spacecraft propulsion due to their high specific impulse and therefore efficiency with respect to propellant use, the mass of propellant which can be carried by the spacecraft without depreciating other subsystems still limits the lifetime of the mission.

\begin{figure} 
	\centering
	\includegraphics[width=\linewidth]{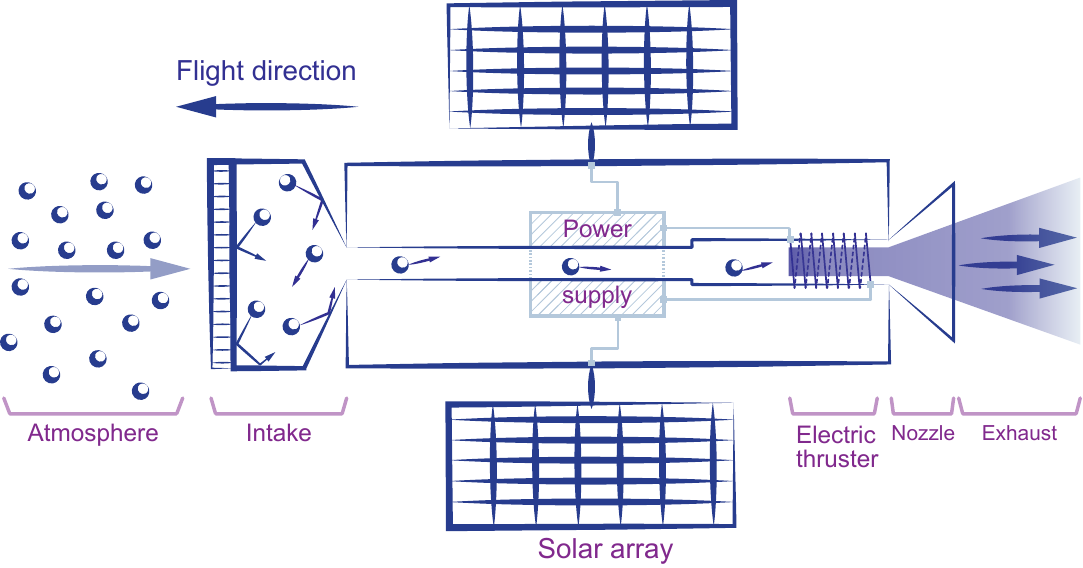}
	\caption{Generalised concept of an atmosphere-breathing electric propulsion (ABEP) system.}
	\label{F:ABEP}
	\end{figure}

Atmosphere-breathing electric propulsion (ABEP) systems on the other hand, shown in \cref{F:ABEP}, propose to collect the oncoming atmospheric gas flow and to use this as the propellant for an electric thruster \cite{Nishiyama2003,DiCara2007}. Using this principle, the lifetime of the spacecraft can be significantly extended beyond current designs as the need for on-board propellant storage eliminated.

The removal of propellant storage offers the opportunity to reduce the spacecraft mass and therefore the launch mass. However, this is dependent on the mass and efficiency of the intake/compressor unit, thruster assembly, and any additional power-raising systems which are required, for example additional deployable solar arrays \cite{Romano2018c}. 

The drag contribution of these additional components also requires consideration as the corresponding thrust requirement will also increase accordingly. For example, analytical approaches to analytical intake design \cite{Romano2016,Binder2016} show that the efficiency of intakes is expected to reduce with an increasing ratio between the intake collection area and thruster inlet area, a result which has been verified against concept intake designs from JAXA \cite{Hisamoto2012} and BUSEK \cite{Hohman2012a}. This means that for a fixed thruster inlet area, as the required thruster mass flow rate increases (for example with reducing altitude and increasing atmospheric density) the collection area needs to increase more rapidly. This has further implications on the magnitude of the drag force which requires compensation.

An optimal altitude range for ABEP propulsion therefore exists for given system performance, above which the atmospheric flow is too rarefied to provide a sufficient mass flow rate of propellant and conventional electric propulsion may be able to provide a reasonable lifetime, and below which the required thrust and available on-board power becomes prohibitive. A review of current ABEP system concepts, rarefied atmospheric intake design, and electric thruster development, is provided by \citet{Schonherr2015}.

The abundance of atomic oxygen in VLEO presents a further complication to the implementation of ABEP as erosion of accelerating grids, electrodes, and discharge channels will result in degraded thruster performance over time. However, concepts for contactless thrusters are current under development, for example an inductive plasma thruster \cite{Romano2018c} which has no electrodes immersed in the plasma and is therefore more resilient to this erosion.

\section{Earth Observation in Very Low Earth Orbits}
\label{S:EO}

Whilst many of the positive features of VLEO described previously are applicable to space missions of all types, Earth observation missions in particular may be significantly benefited. In the following sections, the performance of different EO systems with decreasing is analysed and explored.
 
\subsection{Optical Systems}
Optical systems can be generally classified into three primary categories:
\begin{enumerate}[label=\roman*.]
\item Panchromatic: imagery sensitive to a broad range of wavelengths of visible light, generally represented in black and white or grayscale.
\item Multispectral: imaging in a small number of discrete spectral bands (small ranges of wavelengths). At minimum, the visible spectrum of red, green, and blue light is represented, but depending on the application many bands can be can captured including infrared and ultraviolet spectra.
\item Hyperspectral: imagery is collected in many (up to hundreds or thousands) of narrow and contiguous spectral bands.
\end{enumerate}

\begin{figure}
	\centering
	\includegraphics[width=90mm]{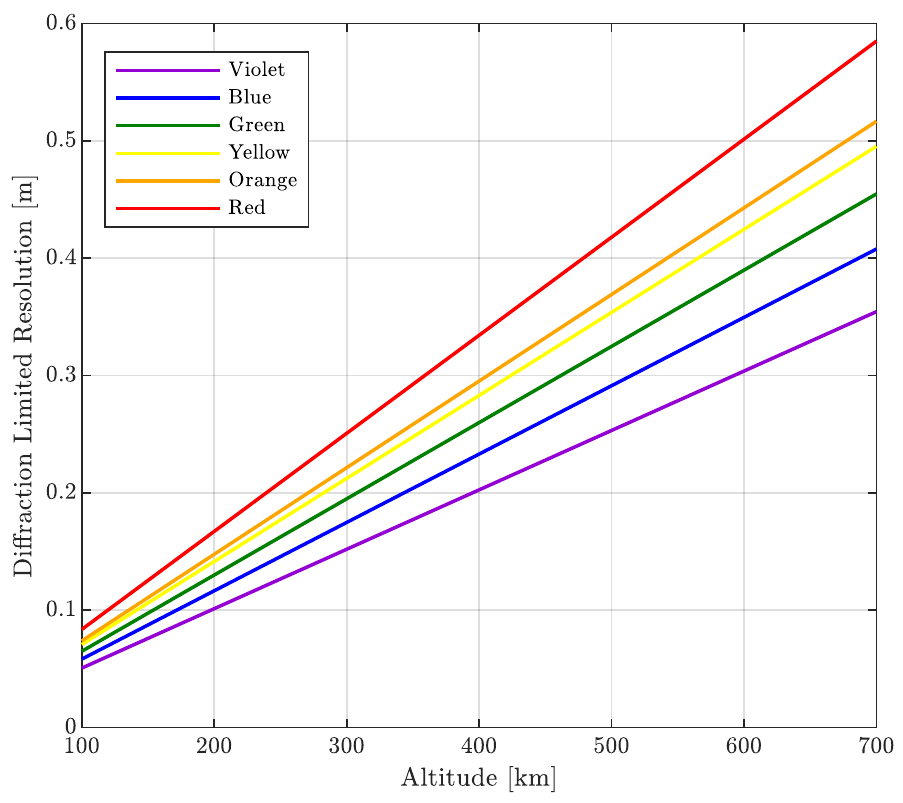}
	\caption{Relationship between altitude and diffraction limited ground resolution for visible light of different wavelengths (for a nadir-pointing aperture with a nominal diameter of \SI{1}{\meter}).}
	\label{F:SpatialRes}
	\end{figure}

For a nadir-pointing telescope, the relationship between diffraction limited resolution and altitude for different wavelengths of light and a fixed lens diameter is given by \cref{E:GRD} and shown in \cref{F:SpatialRes}. A reduction in altitude by \SI{50}{\%} results in an improvement in diffraction limited resolution of a factor of 2 (ie. half the Ground Resolution Distance). Using the same relationship, the aperture diameter can proportionally be reduced with the orbital altitude, demonstrating a benefit in payload sizing which can be achieved whilst maintaining the same diffraction limited resolution.

\begin{figure}
	\centering
	\includegraphics[width=90mm]{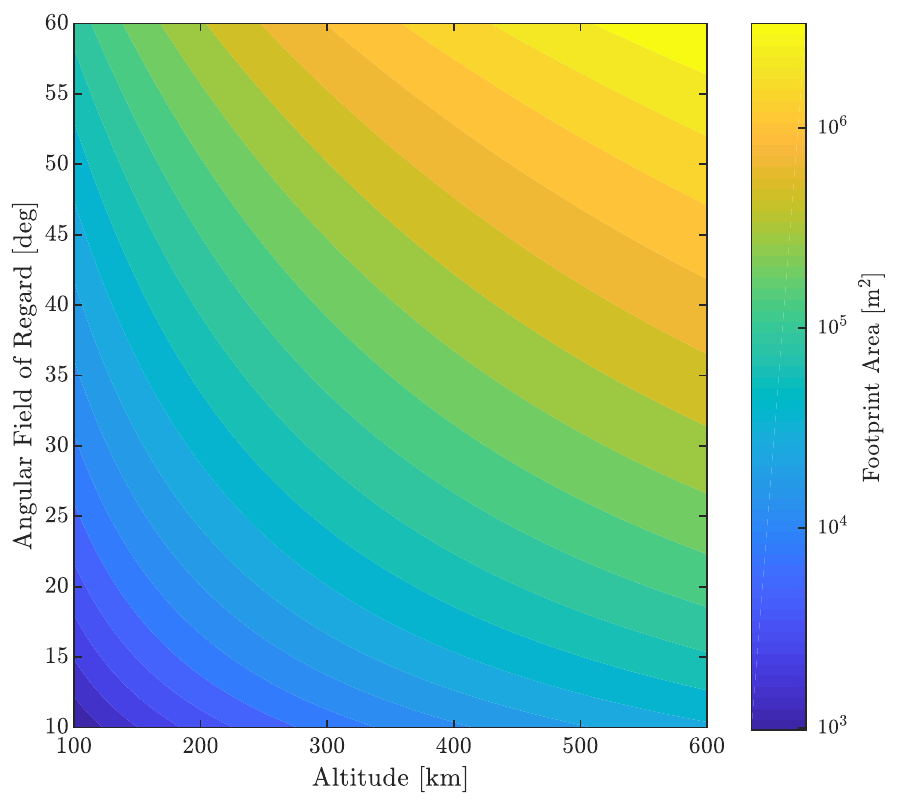}
	\caption{Variation in footprint area with altitude and angular field of regard.}
	\label{F:Area_FoR_Alt}
	\end{figure}

\begin{figure}
	\centering
	\includegraphics[width=90mm]{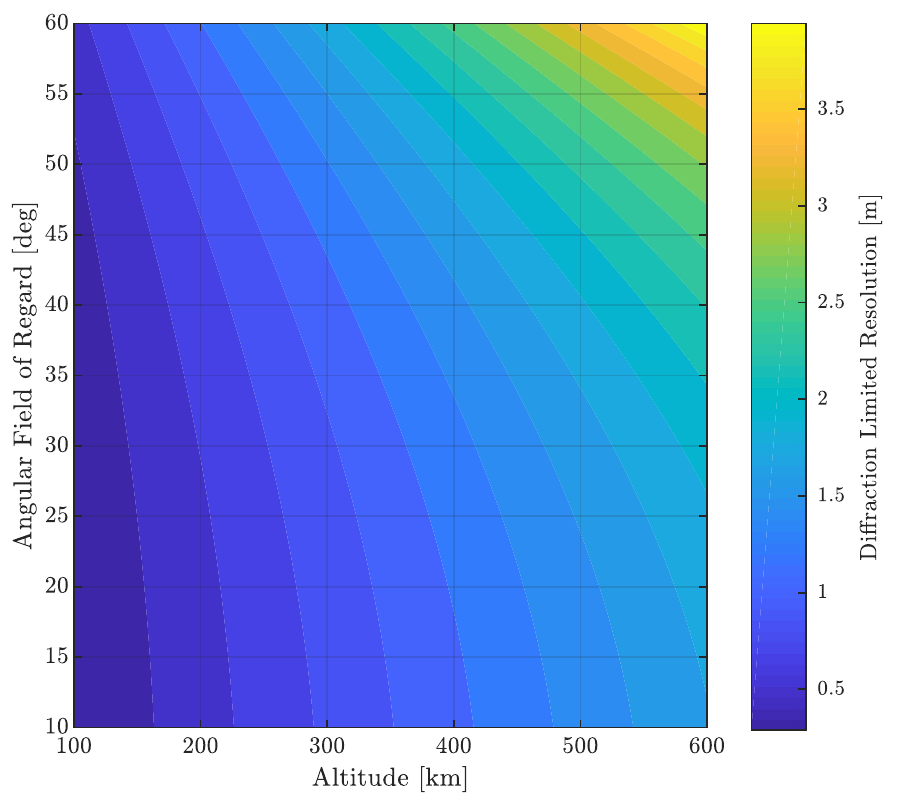}
	\caption{Variation in diffraction limited resolution with spacecraft off-nadir pointing angle and altitude for light at \SI{700}{\nano\meter} wavelength.}
	\label{F:GSD_FoR_Alt}
	\end{figure}

As the altitude of the spacecraft is reduced, referring to the geometry in \cref{F:FootprintGeometry} and \cref{E:FootprintArea}, the total footprint area available to the spacecraft for a given angular field of regard will decrease as demonstrated in \cref{F:Area_FoR_Alt}. For a given altitude, as a result of the longer distance to the edge of the available footprint area, the resolution achievable with increasing field of regard will also decrease. The effect of this is greater at higher-altitudes, demonstrated in \cref{F:GSD_FoR_Alt}.

If high off-nadir pointing performance is considered the resolution performance can vary significantly across the footprint. The elevation angle $\varepsilon$ at the edge of the sensor footprint should also be considered as features at the target may become highly distorted or obscured at low angles of elevation.

Whilst the total MTF of an optical system does not demonstrate a strong dependence on orbital altitude (discussed in \cref{S:MTF}), some secondary effects or system design trade-offs should be considered:
\begin{enumerate}[label=\roman*.]
\item The optical contribution to MTF is generally improved by using larger optical apertures or shorter focal length \cite{Boreman1995} and is therefore dependent on the payload size, design, and specification.
\item Whilst the atmospheric contribution to MTF is not directly related to orbital altitude, the range through the lower atmosphere which an image is acquired can affect the quality and will therefore vary with the off-nadir pointing angle utilized by the system.
\item MTF contributions from platform vibrations can be significantly influenced by the structural design and environmental factors. Density fluctuations, thermospheric wind effects, and the associated aerodynamic interactions may therefore influence this contribution to the MTF and will vary with the operational altitude and environmental conditions.
\end{enumerate}

With regards to radiometric performance for optical based systems, a reduction in altitude will either enable smaller diameter apertures to be used whilst maintaining a given SNR. Alternatively, for a given sensor and aperture the dynamic range can be increased and SNR improved.

The benefits in spatial and radiometric performance with reducing altitude are particularly pertinent for hyperspectral instruments which are typically radiometrically and therefore also spatially constrained due to the narrow width of the individual imaging bands and therefore low SNR \cite{Transon2018}.

Low-cost panchromatic and multispectral optical imaging platforms for both coverage and high-resolution applications will also benefit from lower available orbital altitudes as the spatial resolution and SNR can be improved or smaller diameter optical apertures utilised, thus reducing mass and integration requirements.

\subsection{Passive Infrared and Radar}
Passive infrared and radar (radiometer) payloads sense either reflected or emitted radiation from the Earth. Applications for Earth orbiting systems include thermal infrared observation, microwave imaging, and global navigation satellite system (GNSS) reflectometry (principally for sea-state and wind-speed monitoring) and radio occultation (for atmospheric state and composition) \cite{Pelton2013,Poghosyan2017,Selva2012}.

Like optical observations, these methods are passive and therefore similarly benefit from reduction in the orbital altitude through improved spatial resolution and radiometric performance.

\subsection{Real-Aperture Radar}
Real aperture radar or side-looking airborne radar (SLAR) devices can be used for altimetry or scatterometer applications and offer the ability to penetrate cloud cover or distinguish objects by surface texture or roughness.

These devices are also constrained by the Rayleigh criterion (see \cref{E:GRD}). However, because of the longer-wavelength of radio waves, radar naturally has a larger diffraction limited resolution. As an example, the nadir ground resolution of a \SI{5}{\centi\meter} wavelength radar with a \SI{1}{\meter} aperture diameter and at an orbital altitude of \SI{300}{\kilo\meter} is \SI{18.3}{\kilo\meter}. This demonstrates the limited use of real aperture radar for detection of ground-based features. However, for applications such as altimetry and ocean/wave-height measurement, higher range resolution (vis. height) can be improved by pulse compression methods to provide useful output \cite{Pelton2013,Lacomme2001}.
%

The angular (ambiguity) resolution of radar $\delta_A$ can also be expressed using the half-power (\SI{-3}{dB}) beamwidth angle $B$ at the range $R$, describing the range at which two equally distant targets can be distinguished from each other \cite{RadarSystemsPanelIEEEAerospaceandElectronicSystemsSociety2017}.
\begin{equation} \delta_A  = 2 R \sin{\frac{B}{2}} \end{equation}

The range (ambiguity) resolution $\delta_R$ of a radar describes the minimum linear distance between two targets along the same path from the antenna at which they can be distinguished from each other. For a pulse (rectangular step) waveform, this can be determined from the pulse-width $\tau$ and the speed of light $c_0$ \cite{RadarSystemsPanelIEEEAerospaceandElectronicSystemsSociety2017}.
\begin{equation} \delta_R = \frac{c_0 \tau}{2} \label{E:RadarRangeRes} \end{equation}

The angular resolution of a real aperture radar system is improved by reducing the range to the target and can therefore generally be improved with a reduction in altitude. However, the range resolution is independent of the distance to the target. The combination of the angular and range resolution can be used to define a resolution cell which describes the spacing required to distinguish between multiple targets.

Due to the active component of a radar, the radiometric performance of these systems differs significantly from optical systems. The radar principle is based on the directed transmission of electromagnetic waves, backscattering by different surfaces and materials, and subsequent collection of the returned signal.

The $P_r$ received signal power can be related to the transmitted signal power $P_t$ and receiving and transmitting antenna gains $G_r$ and $G_t$. The distance (range) to the reflecting target is given by $R$, the radar wavelength $\lambda$, and the backscattering or radar cross-section by $\sigma$ \cite{Raemer1997}.
\begin{equation}
\label{E:Radar}
P_r = \frac{P_t G_t G_r \lambda^2 \sigma}{(4\pi)^3 R^4}
\end{equation}

As the signal for an active radar system must travel both the distance to and back from the target, these systems can significantly benefit from any reduction in altitude, illustrated by the relationship of received power to the inverse of the fourth power of the range to the target. Consequently, the transmitting power required for such a system at a lower altitude can be significantly decreased whilst maintaining a similar signal to noise ratio.

The radiometric resolution for a radar is given by the ability of the detector to distinguish between targets with similar backscatter coefficient against the signal intensity and image speckle. An expression for the radiometric resolution $S_{rd}$ can be given considering the average backscatter coefficient $\sigma_0$ and associated standard deviation $\sigma_p$ \cite{Lacomme2001}.
\begin{equation}
S_{rd} = 10 \log_{10} \left( 1 + \frac{\sigma_p}{\sigma_0}\right)
\end{equation}

Noise sources which contribute to degradation in the radiometric resolution include speckle noise resulting from the interference between backscattered waves, background thermal noise, noise internal to the sensor, and quantisation (analogue-to-digital conversion).

The SNR at the receiver for a real aperture radar can be expressed using the radar equation (\cref{E:Radar}) and utilising the following parameters: Boltzmann's constant $k$; the effective noise temperature $T_e$; the receiver noise bandwidth $B_n$; and the receiver noise factor (or ratio between input and output SNR) $\overline{\mathit{NF}}$ \cite{Jain2013}.
\begin{equation}
\mathit{SNR} = \frac{P_t G_t G_r \lambda^2 \sigma}{(4\pi)^3 k T_e B_n \overline{\mathit{NF}} R^4}
\end{equation}

The result of this expression corresponds to the relationship between range and received power and demonstrates an improvement in SNR with the inverse of the fourth power of range with decreasing altitude. Alternatively, the SNR or transmitter power requirement for a monostatic radar can be shown to be improved with the square of the antenna area $A_r$ and related to the antenna efficiency $\eta_{\mathit{ant}}$.
\begin{equation}
\mathit{SNR} = \frac{P_t \eta_{\mathit{ant}}^2 A_r^2 \sigma}{4\pi k T_e B_n \overline{\mathit{NF}} R^4}
\end{equation}

\subsection{Synthetic Aperture Radar}
Synthetic aperture radar (SAR) is an implementation of radar which allows for significantly improved resolution in the velocity direction of a spacecraft. A SAR payload is a typically side-facing radar which detects both the amplitude and phase of backscattered signal. 

In contrast to real aperture radar, SAR takes advantage of the motion of the vehicle and observes targets over the total duration that they fall within the radar beamwidth. As the vehicle has moved during this period a larger aperture is synthesised and a higher resolution can be achieved. The length of this synthetic aperture can be calculated by considering the platform along-track velocity $V_a$, illumination time $T_i$, and azimuth pointing angle $\varphi$
\begin{equation}
\label{E:SAR_Length}
L_{\mathit{SAR}} = V_a T_i \sin{\varphi}
\end{equation}

\begin{figure} 
	\centering
	\def\svgwidth{\linewidth}
	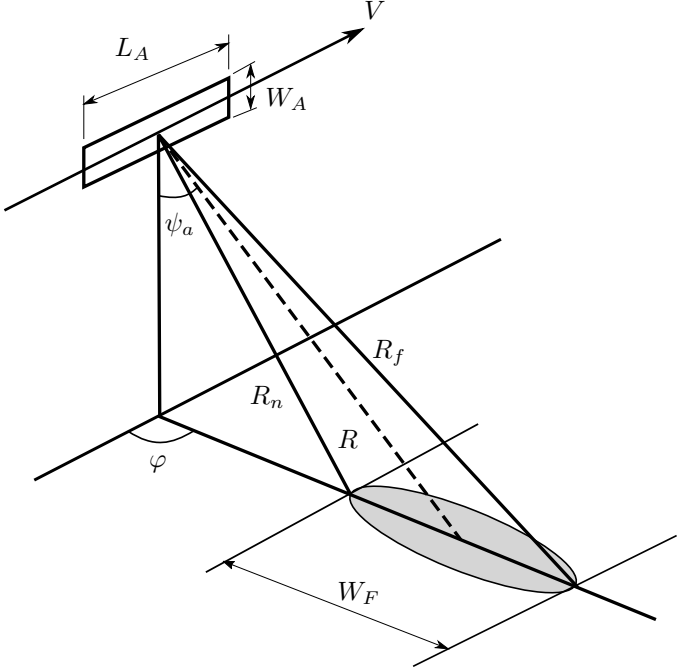
	\caption{SAR footprint geometry. Adapted from \citet{Wertz2011}.}
	\label{F:SARgeometry}
	\end{figure}

For the simplest case, a side-looking SAR sensor (indicated in \cref{F:SARgeometry}), where $\varphi=\ang{90}$, the maximum azimuth (along-track) resolution of a SAR antenna can be defined as function of only the antenna length in that direction $L_A$ \cite{Lacomme2001}.
\begin{equation} \delta_\varphi \ge \frac{L_A}{2} \label{E:SARAziRes} \end{equation}

The azimuth, or along-track resolution $\delta_\varphi$ of SAR is therefore independent of the wavelength, velocity, and range and proportional to the antenna length. Contrastingly to a real aperture radar, and somewhat counter-intuitively, the along-track resolution of SAR improves as the antenna size is reduced.

Similar to the range resolution of a traditional radar (see \cref{E:RadarRangeRes}), the cross-track (range) resolution $\delta_c$ of SAR is dependent on the speed of light $c_0$, pulse-width $\tau$, and for a side-facing antenna can be defined using the angle between the nadir and the slant range to the target $\psi_a$.
\begin{equation}
\delta_c = \frac{c_0 \tau}{2\sin{\psi_a}}
\end{equation}

The cross-track resolution of SAR can therefore be improved by increasing the off-nadir viewing angle and has a theoretical maximum at \ang{90}.

The antenna width $W_A$ is dependent on the wavelength, swath width, range, and incidence, and therefore affects the area of ground which can be covered by the SAR in a pass.
\begin{equation}
W_A = \frac{\lambda R}{W_F \cos{\psi_a}}
\label{E:SAR_Width}
\end{equation}

The minimum antenna area can be seen to increase for greater wavelength, range, and incidence angle. The sizing of a SAR antenna will therefore benefit from a reduction in orbit altitude for the same angle of incidence. However, SAR is still subject to ambiguity constraints based on the frequency of the transmitted pulse, or pulse repetition frequency (PRF). In the along-track direction, the PRF ($1/\tau$) is limited by the velocity and antenna length, such that the vehicle only translates half the length of the antenna during each pulse (azimuth ambiguity) \cite{Tomiyasu1978}. 
\begin{equation}
\mathit{PRF}_{min} = \frac{1}{\tau_{min}} > \frac{2V_a}{L_A}
\end{equation}

Similarly, to avoid detection of multiple echoes in the cross-track direction (range ambiguity), a maximum PRF is defined based on the range to near $R_n$ and far $R_f$ sides of the sensor footprint. 
\begin{equation}
\mathit{PRF}_{max} \le \frac{1}{2\tau_{max} + 2 \left( R_f-R_n \right)c_0^{-1}}
\end{equation}

As the ambiguity constraints on PRF are based on geometric considerations of the antenna, a minimum antenna area can be determined which is dependent on the ratio between the maximum and minimum PRF \cite{Tomiyasu1978}.
\begin{equation}
A_{min} = L_A W_A = \frac{\mathit{PRF}_{max}}{\mathit{PRF}_{min}}\frac{4V_a\lambda R}{c_0}\tan{\psi_a}
\end{equation}
The SNR of a SAR payload is given by \citet{Tomiyasu1978} and \citet{Cutrona1990}.
\begin{equation}
\mathit{SNR} = \frac{P_{av} A_r^2 \eta_{\mathit{ant}}^2 \delta_c \sigma}{8\pi k T_r R^3 \overline{\mathit{NF}} V_a \lambda l_s}
\end{equation}
where $\eta_{\mathit{ant}}$ is an antenna efficiency factor, $\sigma$ is the radar cross-section, $T_r$ is the receiver absolute temperature, $\overline{\mathit{NF}}$ is a relative noise factor, and $l_s$ is the total system loss. The SNR is reduced with wavelength, platform velocity, and the cube of the target range. 

This contrasts with conventional radar which reduces with the fourth-power of the target range. This improvement results from the integration of a number of pulses during the generation of the synthetic aperture. The SNR also improves with lower resolution in the cross-track direction but is independent of the azimuth resolution. If the minimum antenna area is considered (proportional to the range), the SNR relationship with range is also reduced to an inverse square function for a SAR payload. A reduction in orbit altitude whilst maintaining the minimum antenna area will therefore only improve the power requirement linearly.

\begin{figure}
	\centering
	\includegraphics[width=90mm]{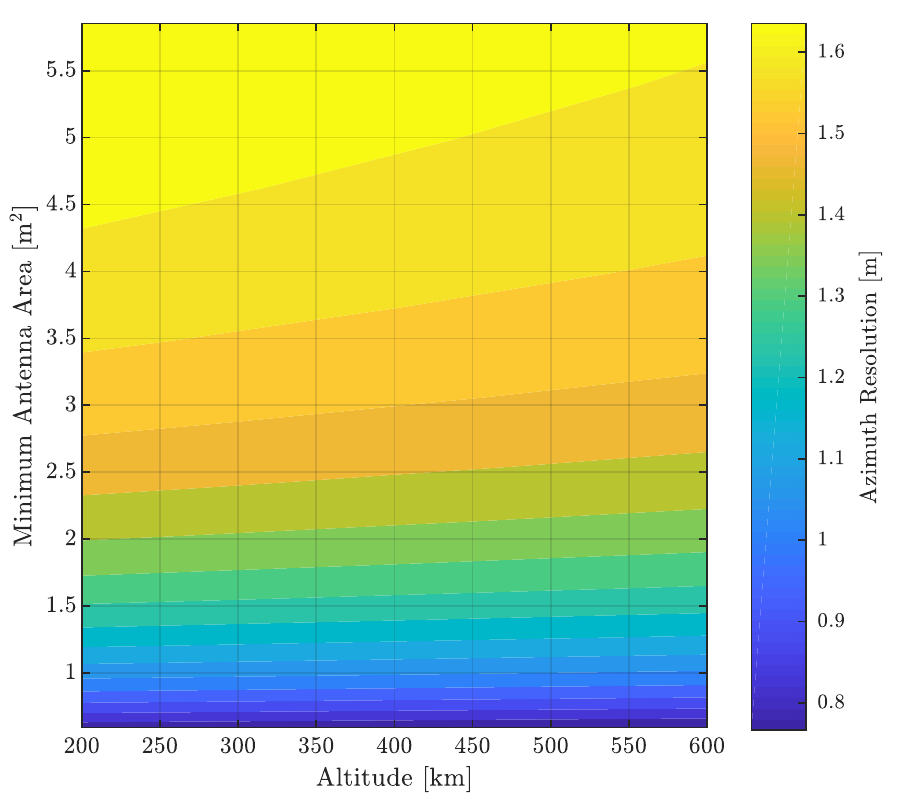}
	\caption{Relationship between SAR minimum antenna area and azimuth resolution for varying orbital altitude, a swath width of \SI{30}{\kilo\meter}, and viewing angle of \ang{45}.}
	\label{F:SAR_Res}
	\end{figure}

\begin{figure} 
	\centering
	\includegraphics[width=90mm]{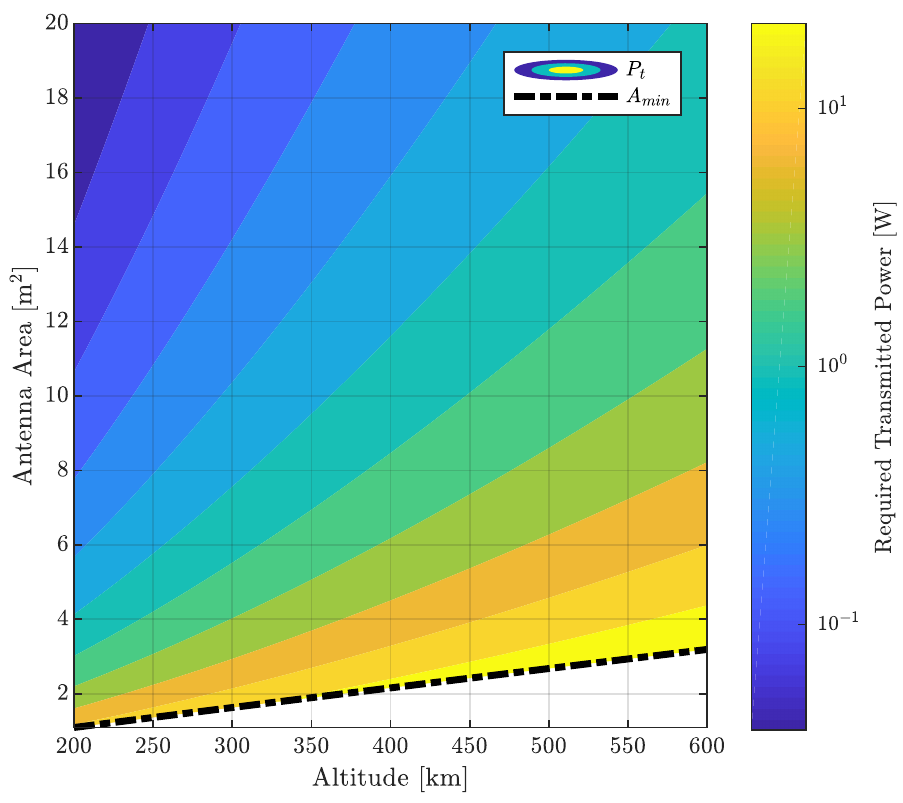}
	\caption{Relationship between SAR power requirement, antenna area and varying orbital altitude for a SNR of \num{3}. A swath width of \SI{30}{\kilo\meter}, viewing angle of \ang{45}, and PRF ratio of \num{1.2} are used.}
	\label{F:SAR_Power}
	\end{figure}

\begin{figure} 
	\centering
	\includegraphics[width=90mm]{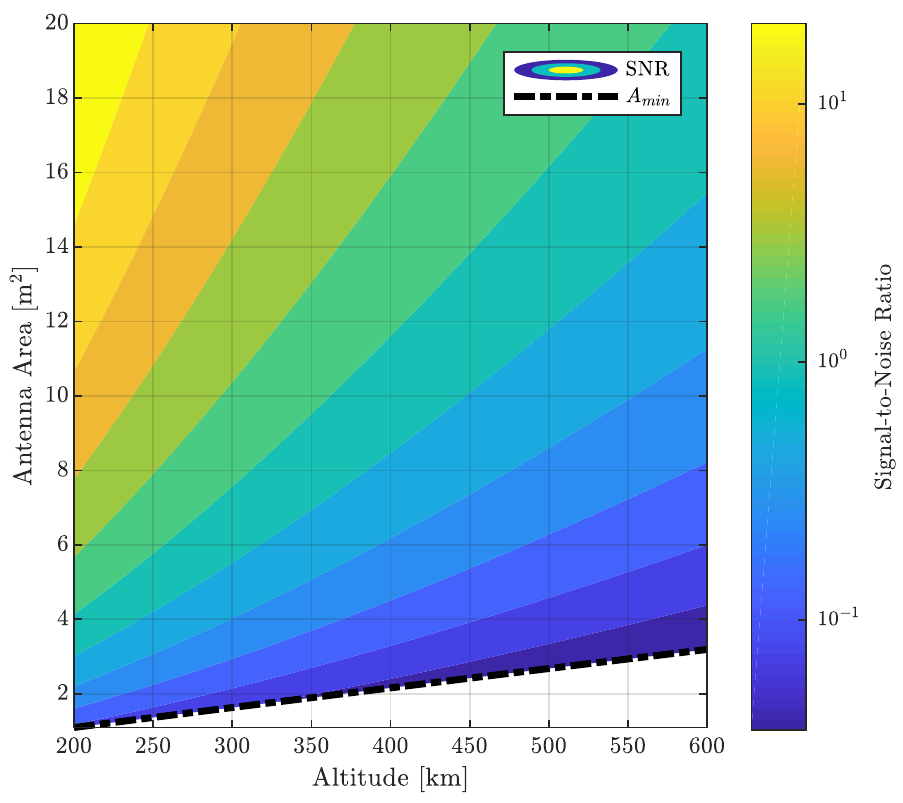}
	\caption{Relationship between SAR signal-to-noise ratio, antenna area, and varying orbital altitude. A power of \SI{5}{\milli\watt}, a swath width of \SI{30}{\kilo\meter}, viewing angle of \ang{45} and PRF ratio of \num{1.2} are used.}
	\label{F:SAR_SNR}
	\end{figure}

Parametric relationships between power, resolution, and antenna area are illustrated in \cref{F:SAR_Res,F:SAR_Power,F:SAR_SNR}. The general independence of SAR resolution with altitude and the inverse relationship with antenna area, is shown in \cref{F:SAR_Res}, demonstrating that the azimuth resolution in fact benefits principally from a smaller physical along-track antenna length (see \cref{E:SARAziRes}).

The relationship between altitude (for a given off-axis viewing angle), minimum antenna area, and power is shown in \cref{F:SAR_Power}. The benefit in transmitter power requirement with a reduction in orbiting altitude is demonstrated. An increase in antenna area is also shown improve the power requirement for a fixed signal to noise level. Similarly, the SNR for a fixed power input is shown in \cref{F:SAR_SNR} to improve for a reducing orbital altitude and increasing antenna area.

The selection of SAR antenna dimensions and total area is therefore a trade-off between the resolution required and the power available on the platform. A further consideration for design of spacecraft for SAR instruments in VLEO is the impact of the antenna and required solar arrays on the aerodynamic performance and drag profile. However, as the antenna can be oriented along the length of the spacecraft with a small cross-sectional area in comparison to optical payloads, the spacecraft configuration can remain quite compact and suitable for use in the high-drag environment of VLEO.

\subsection{Lidar}
Lidar sensors have also been used on spacecraft for meteorological and atmospheric investigation \cite{Pelton2013}, and generation of digital elevation models and terrain mapping activities \cite{Sun2018}. The range resolution or vertical accuracy of a lidar sensor is not dependent on the altitude, but on the available resolution of the available clock or timing measurement chain. The range of lidar is however dependent on the reflected signal strength, and ambient noise factors. A reduction in altitude or range to target will therefore improve SNR or allow a lower power emission.

Similarly to real aperture radars, the returned power of the signal improves by a power of four with a reduced range to the target. This returned power is also dependent on the beam divergence $\beta$, atmospheric $\eta_{\mathit{atm}}$ and system transmission factors $\eta_{\mathit{sys}}$, and the target backscatter coefficient $\sigma_0$ \cite{Kashani2015}.
\begin{equation}
P_r = \frac{P_t \eta_{\mathit{atm}} \eta_{\mathit{sys}} D_r^2 \sigma_0}{4 \pi R^4 \beta^2}
\end{equation}

The range resolution for lidar, important for altimetry applications, can be calculated in the same way as for radar payloads, given in \cref{E:RadarRangeRes}, and is dependent on the pulse-width of the emitted signal.

The theoretical minimum beam divergence of a lidar instrument determines the size of the footprint which is projected onto the ground, and therefore the smallest features which can be identified \cite{Bufton1989,Pelton2013}. For a Gaussian beam (diffraction limited) the angular divergence $\Theta$ is a function of the wavelength $\lambda$ and the beam width at the focus (waist) $w_0$.
\begin{equation} \beta = \frac{2 \lambda}{\pi w_0} \end{equation}

The projected footprint diameter can subsequently be calculated by considering the instrument pointing direction and range to the target/ground. In comparison to radar, the significantly shorter (typically near-infrared) wavelengths used in lidar result in much smaller ground footprints and can therefore achieve a higher linear resolution. In a line scanning mode the measurement spacing or spatial resolution $\delta_L$ of lidar is dependent on the pulse repetition frequency (PRF) and the ground velocity $V_g$ \cite{Sun2018,Pelton2013}. 
\begin{equation} \delta_L = \frac{V_g}{\mathit{PRF}} \end{equation}

As the altitude is lowered, the orbital velocity will increase, and the spatial resolution due to the pulse frequency will decrease for the same $\mathit{PRF}$. 

\section{Impact and Applications of Very Low Earth Orbits for Earth Observation}
In the preceding sections, very low Earth orbits have been shown to offer a number of advantages over traditional LEO altitudes. For EO applications, these principal benefits are enhanced ground resolution, improved radiometric performance, and improved communications link budgets. Alternatively, a reduction in system cost may be achieved through the development and launch of smaller spacecraft which can provide equivalent capability to that of traditional systems at higher orbital altitudes. These spacecraft could also be deployed in larger numbers and into less traditional orbits \cite{Ouma2016}, forming constellations which can offer more frequent revisit opportunities and therefore improved temporal resolution of imagery or data.

In the downstream markets, both direct and through value-adding services, the availability of larger volumes of lower cost, more timely, or better quality imagery and data products has significant value to both commercial end-users and global societal, sustainability, and environmental objectives \cite{,UN2018,CommitteeonEarthObservatoinSatellites2018}. For example, in boarder security and maritime surveillance higher spatial resolution can facilitate the identification of smaller vehicles and vessels and improve classification \cite{Greidanus2008,Ruddick2008,Wolfinbarger2015}, enabling better assessment of risk and vulnerability. Similarly, for applications such as agriculture, water-security, climate-change, infrastructure monitoring, and location-based services (or asset tracking), enhanced resolution can enable more detailed change-detection \cite{Navalgund2007,Dowman2005}, but also needs to be supported by suitable revisit times and affordable data continuity. Enhancement of radiometric resolution can have similar benefits, some of which have been demonstrated for a range of these applications by the Landsat-8 mission \cite{Roy2014}.

From a humanitarian aspect, initiatives such the International Charter ``Space and Major Disasters'', the Center for Satellite-Based Crisis Information (ZKI), and the Copernicus programme already facilitate access to multi-source EO data products \cite{Voigt2007}. However, demand for higher resolution and more timely imagery is needed to enable a more rapid and precise response \cite{Denis2016}, Future VLEO satellite systems may be able to address this demand supporting improved humanitarian assistance and crisis management.

In the midstream segment, the increase in volume or value of the obtained imagery and data will support growth of the Earth observation market, principally through commercial sales \cite{Probst2017}. In the upstream Earth observation markets, the corresponding demand for VLEO systems will offer opportunities for continued commercial and industrial growth in areas such as spacecraft development, manufacturing, and launch. Operating spacecraft effectively and efficiently in VLEO will also require the development of new technologies, encouraging innovation and disruption in the market.

\section{Conclusions and Recommendations}
Given the commercial and societal benefits which may be realised by operating Earth observation spacecraft in VLEO, there has been a renewed push to revisit the challenges associated with this reduction in orbital altitude. The most significant of these is the increased atmospheric density at lower altitudes which increases aerodynamic drag and therefore reduces orbital lifetime. 

The presence of highly-reactive atomic oxygen, which is typically the most abundant gas species in VLEO, can affect and degrade material performance, in particular sensitive optical sensor surfaces and thermal coatings \cite{Samwel2014,Finckenor2015}. Research towards the identification of materials which are resistant to both erosion by atomic oxygen and can reduce aerodynamic drag is therefore an active area of interest \cite{Roberts2017}. If combined with appropriate spacecraft geometries and platform designs, these materials will also support the exploitation of novel aerodynamic control manoeuvres which can aid sustained operation of spacecraft in lower altitude orbits.

Improved understanding of the atmospheric density and thermospheric winds is also necessary to facilitate  aerodynamic control methods and high-precision operations at lower orbital altitudes. In particular, new measurements of the variation in the lower thermosphere over different time-scales would help to improve  understanding and modelling efforts in this area. New missions with this aim are currently being proposed, for example the new ESA-led Daedalus Earth Explorer satellite \cite{Sarris2019} which is now proceeding to feasibility studies with a planned launch in 2027--2028.

Development of novel propulsive technologies is required to enable the sustained operation of spacecraft at these lower altitudes. In particular, atmosphere-breathing electric propulsion (ABEP) systems offer the potential for significantly extended lifetime by eliminating the need to store propellant on-board whilst providing effective drag-compensation \cite{Schonherr2015}. One such area seeing promising development is the design of electrodeless thrusters which avoid the erosion of critical components due to the prevalence of of oxidising species in the VLEO environment \cite{Romano2018c}. However, challenges associated with system efficiency and integration into platform designs remain key areas requiring research and development.

Despite the range of Earth observation applications which VLEO may be able to enhance, quantitative assessment or studies of these impacts have not yet been carried out. New business models which focus on these different market segments, from the upstream through to downstream, are therefore required to establish the economic and commercial potential of VLEO systems to these application areas. 

Finally, engineering and system modelling efforts are needed to establish the feasibility and viability of VLEO systems which will combine novel technology developments and new operational concepts. Combination of these systems engineering models with business models for different Earth observation applications will enable identification of the most promising concepts for VLEO and define the roadmap for exploitation and implementation.

\begin{table*}[!t]   
	\begin{framed}
			\nomenclature{$A_F$}{Footprint area}
		\nomenclature{$A_r$}{Receiver Area}
		\nomenclature{$A_{\mathit{ref}}$}{Aerodynamic reference area}
		\nomenclature{$B$}{Beamwidth angle}
		\nomenclature{$B_n$}{Receiver noise bandwidth}
		\nomenclature{$C_F$}{Force coefficient}
		\nomenclature{$D$}{Aperture diameter}
		\nomenclature{$E_{m,x}$}{Mapping error}
		\nomenclature{$E_{p,x}$}{Pointing error}
		\nomenclature{$F_a$}{Aerodynamic force}
		\nomenclature{$G$}{Gain}
		\nomenclature{$L_A$}{Antenna length}
		\nomenclature{$L_a$}{Transmission path loss factor}
		\nomenclature{$L_{\mathit{SAR}}$}{Synthetic aperture length}
		\nomenclature{$N$}{Noise}
		\nomenclature{$P$}{Power}
		\nomenclature{$R$}{Range}
		\nomenclature{$R_\phi$}{Earth radius}
		\nomenclature{$T_E$}{Effective noise temperature}
		\nomenclature{$T_i$}{Illumination time}
		\nomenclature{$T_r$}{Receiver noise temperature}
		\nomenclature{$V_E$}{Earth equatorial velocity}
		\nomenclature{$V_a$}{Along-track velocity}
		\nomenclature{$V_g$}{Ground velocity}
		\nomenclature{$V_{\mathit{rel}}$}{Velocity relative to the atmosphere}
		\nomenclature{$W_A$}{Antenna width}
		\nomenclature{$W_F$}{Footprint Width}
		\nomenclature{$\beta$}{Lidar beam divergence}
		\nomenclature{$\delta_A$}{Radar angular resolution}
		\nomenclature{$\delta_A$}{Radar angular resolution}
		\nomenclature{$\delta_L$}{Lidar spatial resolution}
		\nomenclature{$\delta_R$}{Radar range resolution}
		\nomenclature{$\delta_R$}{Radar range resolution}		
		\nomenclature{$\delta_\Theta$}{Angular resolution}
		\nomenclature{$\delta_c$}{SAR cross-track resolution}
		\nomenclature{$\eta_{\mathit{ant}}$}{Antenna efficiency}
		\nomenclature{$\eta_{\mathit{atm}}$}{Atmospheric transmission factor}
		\nomenclature{$\eta_{\mathit{sys}}$}{System transmission factor}
		\nomenclature{$\lambda$}{Wavelength}
		\nomenclature{$\overline{\mathit{NF}}$}{Noise factor}
		\nomenclature{$\phi$}{Latitude}
		\nomenclature{$\psi$}{Angular field of regards}
		\nomenclature{$\sigma$}{Radar cross-section}
		\nomenclature{$\sigma_0$}{Backscatter coefficient}
		\nomenclature{$\sigma_p$}{Standard deviation of backscatter coefficient}
		\nomenclature{$\tau$}{Pulse width}
		\nomenclature{$\theta$}{Earth central angle}
		\nomenclature{$\varepsilon$}{Elevation angle}
		\nomenclature{$\varphi$}{Azimuth angle}
		\nomenclature{$c_0$}{Speed of light}
		\nomenclature{$f$}{Focal length}
		\nomenclature{$h_T$}{Target altitude}
		\nomenclature{$h_\phi$}{Ellipse altitude}
		\nomenclature{$k$}{Boltzmann's constant}
		\nomenclature{$l_{s}$}{Total system loss}
		\nomenclature{$r_s$}{Orbit (satellite) radius}
		\nomenclature{$x$}{Pixel size}

	\printnomenclature
	\end{framed}
	\end{table*}

\section*{Acknowledgements}
This project has received funding from the European Union's Horizon 2020 research and innovation programme under grant agreement No 737183. This publication reflects only the view of the authors. The European Commission is not responsible for any use that may be made of the information it contains.

\bibliographystyle{elsarticle-num-names}
\bibliography{Benefits_of_VLEO}

\end{document}